\documentclass[preprint,showpacs,amsmath,amssymb,aps,prd,nofootinbib]{revtex4}
\usepackage{epsfig,color,appendix,multirow,ulem}

\newcommand{\diag}{\mathop{\rm diag}}
\renewcommand{\Re}{\mathop{\rm Re}}

\begin{document}

\title{
Towards a realistic model of quarks and leptons,\\ leptonic CP violation and neutrinoless $\beta\beta$-decay}

\author{Y. H. Ahn}
\affiliation{School of Physics, KIAS, Seoul 130-722, Korea}
\email{yhahn@kias.re.kr}

\author{Paolo Gondolo}
\affiliation{Department of Physics and Astronomy, University of Utah,\\
115 South 1400 East \#201, Salt Lake City, UT 84112, USA}
\email{paolo.gondolo@utah.edu}


\begin{abstract}
\noindent In order to explain the fermion masses and mixings naturally, we introduce a specific flavor symmetry and mass suppression pattern that constrain the flavor structure of the fermion Yukawa couplings. Our model describes why the hierarchy of neutrino masses is milder than the hierarchy of charged fermion masses in terms of successive powers of flavon fields. We investigate CP violation and neutrinoless double beta ($0\nu\beta\beta$) decay, and show how they can be predicted and constrained in our model by present and upcoming experimental data. Our model predicts that the atmospheric neutrino mixing angle $\theta_{23}$ should be within $\sim1^{\circ}$ of $45^\circ$ for  the normal neutrino mass ordering (NO), and between $\sim4^\circ$ and $\sim8^\circ$ degrees away from $45^\circ$ (in either direction) for the inverted neutrino mass ordering (IO). For both NO and IO, our model predicts that a $0\nu\beta\beta$ Majorana mass in the limited range $0.035~\text{eV}<|m_{ee}|\lesssim0.15$ eV, which can be tested in current experiments. Moreover, our model can successfully accommodate flavorless leptogenesis as the mechanism to generate the baryon asymmetry in the Universe, provided the neutrino mass ordering is normal, $|m_{ee}|\simeq0.072\pm0.012$ eV, and either $\theta_{23}\simeq44^{\circ}$  and the Dirac CP-violating phase $\delta_{CP}\simeq20^{\circ}$ or $60^{\circ}$, or $\theta_{23}\simeq46^{\circ}$ and $\delta_{CP} \simeq205^{\circ}$ or $245^{\circ}$.
\end{abstract}

\maketitle %
\section{Introduction}
An outstanding puzzle in the standard model (SM) of particle physics is the pattern of fermion masses and mixings. The fermion masses cover a range of at least 12 orders of magnitude. The neutrino mass is bounded by $\sum m_{\nu}\lesssim0.23$ eV (Planck-I) or $\lesssim0.66$ eV (Planck-II)~\cite{Ade:2013zuv}, which is to be compared to the top quark mass $m_{t}\simeq173$ GeV~\cite{PDG}. The mass ratio between the heaviest and the lightest quark (the top and the up quark) is $m_{t}/m_{u}\sim10^{5}$, the mass ratio between the heaviest and the lightest charged lepton (the tau and the electron) is $m_{\tau}/m_{e}\sim10^{3}$, and the mass ratio between neutrinos seems to be $\sim10^2$. Fermion mixing angles follow a different pattern for quarks and leptons: one large and two small mixing angles  for the quarks ($\sim\!13^\circ$, $\sim\!2^\circ$, $\sim\!0.2^\circ$) and a large CP-violating phase ($\sim\!60^\circ$); two large and one small mixing angle for the leptons ($\sim\!33^\circ$, $\sim\!45^\circ$, $\sim\!8^\circ$) and no experimental information yet on the leptonic CP-violating phases.

It is believed that an understanding of the observed pattern of fermion masses and mixings may provide a crucial clue to physics beyond the SM. The two large lepton mixing angles may be telling us about new symmetries not present in the quark sector and may provide a clue to the nature of the quark-lepton physics beyond the SM.
Actually, in the absence of flavor symmetries, particle masses and mixings are generally undetermined in a gauge theory. With a single Higgs in the SM one cannot explain the strong hierarchies in the quark and lepton masses. Of course, one can imagine that the fermion masses and mixings are independent parameters in the SM. However, one cannot calculate them from a fundamental theory. It is natural to suppose that the extreme smallness of the neutrino masses in comparison to the charged fermion masses is related to the existence of a new fundamental scale, and thus to new physics beyond the SM. Large ratios between the masses of fermions of successive generations may be due to suppressions by different powers of the new scale, and there could be a hierarchy in which the masses of the lighter fermions are suppressed by powers of a large new scale (e.g., the seesaw mechanism of~\cite{Minkowski:1977sc} or the Froggatt-Nielsen mechanism of~\cite{Froggatt:1978nt}). A new large scale may also be used to explain why the hierarchy of neutrino masses is milder than the hierarchies of quarks and charged leptons.

In this paper, we introduce a specific flavor symmetry and mass suppression pattern that constrain the flavor structure of the fermion Yukawa couplings and leads to predictions for the fermion masses and mixings. The large fermion mixing angles can be understood by introducing a non-Abelian discrete flavor symmetry group, and the small fermion mixing angles can arise from a mismatch between the residual symmetry of the flavor group  after the discrete flavor symmetry is spontaneously broken.
The mass hierarchies of the fermion sector can be understood by introducing an anomalous $U(1)_X$ global symmetry, in which gauge singlet flavon fields ${\cal F}_i$ couple to dimension-3 or -4 fermion operators with different powers of ${\cal F}_i$. Schematically, the electroweak-scale fermion Lagrangian depends on the flavon fields as
 \begin{eqnarray}
c_0 \, {\cal O}_{0} + c'_1 \, {\cal O}'_{1}\,{\cal F} +
c_1 \, {\cal O}_{1}\,\left(\frac{{\cal F}}{\Lambda}\right)+ c_2 \, {\cal O}_{2}\,\left(\frac{{\cal F}}{\Lambda}\right)^{2} + c_3 \, {\cal O}_{3}\,\left(\frac{{\cal F}}{\Lambda}\right)^{3}+\cdots,
 \label{flavon}
 \end{eqnarray}
where the ${\cal O}'_1$ and the ${\cal O}_{i}$ are dimension-3 and dimension-4 fermion operators, and the coefficients $c'_1$ and $c_i$ are of order 1. Here $\Lambda$ is the scale of flavor dynamics, and the mass scale of the Froggart-Nielsen heavy fields that are integrated out. Since the Yukawa couplings are eventually responsible for the fermion masses they must be related in a very simple way at a large scale in order for intermediate scale physics to produce all the interesting structure in the fermion mass matrices.

We propose a realistic model for quarks and leptons based on an $A_{4}\times U(1)_X$ flavor symmetry~\footnote{It is different from previous works using $A_{4}$ symmetries~\cite{A4} in that the Dirac neutrino Yukawa coupling
constants do not all have the same magnitude.} in the seesaw framework. The seesaw mechanism, besides explaining of smallness of the measured neutrino masses,  has the additional appealing feature of being able to generate the observed baryon asymmetry of the Universe through leptogenesis~\cite{review}. In such a framework the Yukawa couplings are functions of flavon fields which are responsible for making right-handed neutrinos very heavy.

The main theoretical goal of our work is twofold. First, we are going to explain the large and small mixing angles in the lepton and quark sectors, and the enormously various hierarchies spanned by the fermion masses, in terms of successive powers of the flavon field, describing also why the hierarchy of light neutrino masses is relatively mild, while the hierarchy of the charged fermions is strong. Second, we investigate CP violation and neutrinoless double beta ($0\nu\beta\beta$) decay in the lepton sector and show how CP phases and/or $0\nu\beta\beta$-decay can be predicted and/or constrained by the model and/or the present experimental data. Moreover, in our model, since the Dirac neutrino Yukawa couplings are of order 1, a successful explanation of the baryon asymmetry of the Universe through leptogenesis may be possible if the leptogenesis scale is $\sim10^{12}$ GeV, which is below the grand unification scale of $\sim\!10^{16}$ GeV. Implementing such leptogenesis can provide information or constraints on the Dirac CP-violating phase and $0\nu\beta\beta$-decay.

This paper is organized as follows. In the next section, first we lay down the particle content
and the field representations under the $A_4$  flavor symmetry, then we construct Higgs and Yukawa Lagrangians, and finally add a flavor symmetry $U(1)_X$ to build an effective model. In Sec.~III, we discuss how the hierarchies of masses and  mixings in the quark and lepton sectors can be realized after spontaneous symmetry breaking of the $A_4$ flavor symmetry.
In Sec.~IV, we consider leptonic CP violation, $0\nu\beta\beta$-decay and leptogenesis, and we perform a numerical analysis of our model using neutrino oscillation data.
We give our conclusions in Sec.~V.

\section{The Model}

In order to understand the small lepton mixing angle $\theta_{13}\sim8^\circ$ and the two large lepton mixing angles ($\theta_{12}\sim33^\circ, \theta_{23}\sim45^\circ$) as well as the Cabibbo quark mixing angle $\theta_{C}\sim13^\circ$ and the two small quark mixing angles, we propose a model based on an $A_{4}$ flavor symmetry for leptons and quarks, which is an extension of that in Ref.~\cite{Ahn:2012cg}. In addition, in order to describe the strong hierarchy of charged fermion masses and the mild hierarchy of neutrino masses, we use the mechanism in Eq.~(\ref{flavon}), imposing a continuous global $U(1)_X$ symmetry under which the fermions are distinguished.\footnote{Since Goldstone modes resulting from spontaneous $U(1)_X$ symmetry breaking are not phenomenologically allowed, $U(1)_X$  is explicitly broken by a soft-breaking term.} Finally, to enforce that only the Higgs field $\eta$ and not $\Phi$ contributes to up-type quark and charged-lepton mass terms, we have introduced a discrete $Z_{2}$ symmetry. Mathematical details of the group $A_{4}$ are given in Appendix~\ref{A4group}.

We extend the standard model (SM) by the inclusion of right-handed neutrinos and additional Higgs fields. The field content of our model and the field assignments to $A_{4} \times SU(2)_L\times U(1)_Y\times U(1)_X\times Z_{2}$ representations are summarized in Table~\ref{reps}, which we now describe (the $U(1)_X$ assignments are explained in Section \ref{sec:yukawa} below).

The left-handed lepton doublets
\begin{eqnarray}
L_e=\begin{pmatrix} \nu_{Le} \\ e_{L} \end{pmatrix},
\quad
L_\mu=\begin{pmatrix} \nu_{L\mu} \\ \mu_{L} \end{pmatrix},
\quad
L_\tau=\begin{pmatrix} \nu_{L\tau} \\ \tau_{L} \end{pmatrix},
\end{eqnarray}
are respectively assigned to the $\mathbf{1}$, $\mathbf{1^{\prime}}$, $\mathbf{1^{\prime\prime}}$ representations of $A_{4}$. That is, they are $S$-flavor-even and have $T$-flavor $0$, $+1$, and $-1$, respectively. The right-handed charged leptons
\begin{align}
e_R, \qquad \mu_R, \qquad \tau_R ,
\end{align}
are also assigned to the $\mathbf{1}$, $\mathbf{1^{\prime}}$, $\mathbf{1^{\prime\prime}}$ representations of $A_{4}$, respectively. They have thus the same $S$-flavor-parity and $T$-flavor of the left-handed charged lepton in the same generation. In other words, electrons and electron-neutrinos have $T$-flavor 0, muons and muon-neutrinos have $T$-flavor $+1$, and tau and tau-neutrinos have $T$-flavor $-1$. The right-handed neutrinos
\begin{align}
\label{N_R}
N_R = \begin{pmatrix} N_{R1} & N_{R2} & N_{R3} \end{pmatrix}
\end{align}
are a triplet of $A_4$ (i.e., are in the $\mathbf{3}$ representation of $A_4$). They can either be written in the $S$-diagonal matrix representation as in Eq.~(\ref{N_R}), where $N_{R1}$ is $S$-flavor-even and $N_{R2}$ and $N_{R3}$ are $S$-flavor-odd, or in the $T$-diagonal representation
\begin{align}
N_R =  \begin{pmatrix} N_{R,0} & N_{R,+1} & N_{R,-1} \end{pmatrix} \, U_{\omega}^T,
\end{align}
where $N_{R,t}$ has $T$-flavor $t$ (see Appendix A).

\begin{table}[t]
\caption{\label{reps} Representations of the fields under $A_4 \times SU(2)_L \times U(1)_Y\times U(1)_X\times Z_2$. }
\begin{ruledtabular}
\def\extraspace{\hspace{1em}}
\begin{tabular}{c|ccc|ccc|cc|cc}
\multirow{2}*{Field}
& \multicolumn{3}{c|}{Leptons}
& \multicolumn{3}{c|}{Quarks}
& \multicolumn{2}{l|}{Higgses}
& \multicolumn{2}{l}{Flavons}
\\
& $L_{e}$, $L_{\mu}$, $L_{\tau}$ & $e_R$, $\mu_R$, $\tau_R$ & $N_{R}$ \extraspace
& $Q_{L_1}$, $Q_{L_2}$, $Q_{L_3}$ & $u_R$, $c_R$, $t_R$ & $D_{R}$ \extraspace
& $\Phi$ & $\eta$\extraspace
& $\chi$& $\Theta$ \extraspace
\\
\hline
$A_4$
& $\mathbf{1}$, $\mathbf{1^{\prime}}$, $\mathbf{1^{\prime\prime}}$ & $\mathbf{1}$, $\mathbf{1^{\prime}}$, $\mathbf{1^{\prime\prime}}$ & $\mathbf{3}$ \extraspace
& $\mathbf{1}$  & $\mathbf{1}$ & $\mathbf{3}$ \extraspace
& $\mathbf{3}$ & $\mathbf{1}$ \extraspace
 & $\mathbf{3}$& $\mathbf{1}$ \extraspace
\\
$SU(2)_L$
& $2$ & $1$ & $1$ \extraspace
& $2$ & $1$ & $1$ \extraspace
& $2$ & $2$\extraspace
& $1$& $1$ \extraspace
\\
$U(1)_Y$
& $-1$ & $-2$ & $0$ \extraspace
& $\frac{1}{3} $ & $\frac{4}{3}$ & $-\frac{2}{3}$ \extraspace
& $1$ & $1$\extraspace
& $0$ & $0$ \extraspace
\\
$U(1)_X$
& $-p$ & $7p$, $4p$, $2p$ & $0$ \extraspace
& $q-3p$, $q-2p$, $q$ ~&~ $q+7p$, $q+4p$, $q$ ~&~ $3p+q$ \extraspace
& $0$ & $0$ \extraspace
& $p$& $-p$ \extraspace
\\
$Z_2$
& $+$ & $-$ & $+$ \extraspace
& $+$ & $-$ & $+$ \extraspace
& $+$ & $-$ \extraspace
& $+$& $+$ \extraspace
\\
\end{tabular}
\end{ruledtabular}
\end{table}

We assign the left-handed quark doublets
\begin{eqnarray}
Q_{L_1}=\begin{pmatrix} u_{L} \\ d_{L} \end{pmatrix},
\quad
Q_{L_2}=\begin{pmatrix} c_{L} \\ s_{L} \end{pmatrix},
\quad
Q_{L_3}=\begin{pmatrix} t_{L} \\ b_{L} \end{pmatrix},
\end{eqnarray}
to the $\mathbf{1}$ representation of $A_{4}$. That is, they are all $S$-flavor-even and have $T$-flavor $0$.
The right-handed down-type quarks are assigned to the $\mathbf{3}$ representation of $A_4$, i.e., they are an $A_4$ triplet. They can be written in the $S$-diagonal or in the $T$-diagonal bases as
\begin{eqnarray}
D_{R} = \begin{pmatrix} d_{R} & s_{R} & b_{R} \end{pmatrix} = \begin{pmatrix} d_{R,0~} & d_{R,+1} & d_{R,-1} \end{pmatrix} U_\omega^T.
 \end{eqnarray}
Here $d_{R}$ is $S$-flavor-even, $s_{R}$ and $b_{R}$ are $S$-flavor-odd, and $d_{R,t}$ has $T$-flavor equal to $t$.
Notice the mismatch between the $T$-flavors of right-handed and left-handed down-type quarks.
The right-handed up-type quarks
\begin{align}
u_R, \qquad c_R, \qquad t_R ,
\end{align}
are assigned to the same $A_4$ representation as the left-handed up-type quarks of the same name.

The Higgs sector contains two sets of Higgs fields, according to the order of magnitude of their vacuum expectation value (VEV) after symmetry breaking. Higgs bosons in the first set have VEVs of the order of the electroweak symmetry breaking scale ($\sim 10^2$ GeV). Higgs bosons in the second set have much larger VEVs, and are flavon fields.

The electroweak Higgs fields are an $A_4$ triplet $\Phi$ (in the $\mathbf{3}$ representation) and an $A_4$ singlet $\eta$ (in the $\mathbf{1}$ representation); both are $SU(2)_L$ doublets:
\begin{align}
\Phi & = \begin{pmatrix} \varphi^{+}_1 & \varphi^{+}_2 & \varphi^{+}_3 \\ \varphi^{0}_1 & \varphi^{0}_2 & \varphi^{0}_3 \end{pmatrix} = \begin{pmatrix} \varphi^{+}_{0} & \varphi^{+}_{+1} & \varphi^{+}_{-1} \\ \varphi^{0}_{0} & \varphi^{0}_{+1} & \varphi^{0}_{-1} \end{pmatrix} U_\omega^T ,
\\
\eta & =
\begin{pmatrix} \eta^{+} \\ \eta^{0} \end{pmatrix}.
\label{Higgs}
\end{align}
The fields $\varphi^{+}_j$ and $\eta^{+}$ ($\varphi^{0}_j$ and $\eta^{0}$, resp.) have electric charge $+1$ (0, resp.). The fields $\varphi^{+}_1$, $\varphi^{0}_1$, $\eta^{+}$ and $\eta^{0}$ are $S$-flavor-even, while $\varphi^{+,0}_{2,3}$ are $S$-flavor-odd. The fields $\varphi^{+,0}_{0}$, $\varphi^{+,0}_{+1}$, and $\varphi^{+,0}_{-1}$ have $T$-flavor 0, $+1$, and $-1$, respectively, while $\eta^{+,0}$ have $T$-flavor zero.

The flavon fields are an $A_4$ triplet $\chi$ (in the $\mathbf{3}$ representation) and an $A_4$ singlet $\Theta$ (in the $\mathbf{1}$ representation); both are $SU(2)_L$ singlets:
\begin{eqnarray}
\chi = (\chi_1, \chi_2, \chi_3),
\qquad
\Theta .
\label{Flavon}
\end{eqnarray}

The Higgs doublet $\Phi$, the Higgs singlets $\chi$ and $\Theta$, and the singlet neutrinos $N_R$ are assumed to be triplets under $A_4$, and can so be used to introduce lepton-flavor violation in an $A_4$ symmetric Lagrangian. In our Lagrangian, acquiring non-zero VEVs $\langle\Theta\rangle$ and $\langle\chi\rangle$ breaks the flavor symmetry and the $U(1)_X$ symmetry. The breaking of the $U(1)_X$ symmetry is communicated to the fermions with different powers of the flavon fields $\chi$ and $\Theta$.

\subsection{Low energy Yukawa terms}
\label{sec:yukawa}
We start by designing a concrete model that will induce the desired effective Yukawa Lagrangian in the way of Eq.~(\ref{flavon}). Here we consider only the Lagrangian terms that give rise to lepton and quark masses.

The flavon gauge singlets $\Theta$ and $\chi$ are dynamical at a very high energy scale (namely, the seesaw scale, or the grand unification theory scale). Their VEVs are communicated to the charged fermions through Yukawa couplings and give rise to the fermion masses.
We focus on the particularly interesting possibility that the hierarchical pattern of charged fermion masses can be explained by powers of $\langle{\cal F}\rangle/\Lambda$ according to appropriate flavor symmetries. Since the Yukawa couplings are ultimately responsible for the fermion masses which reflect enormously various hierarchies they must be understood in a very reasonable way: an anomalous $U(1)_X$ global symmetry prevents the direct Yukawa coupling of the SM Higgs doublet to the light fermions. In addition to this,
to obtain a realistic Cabibbo-Kobayashi-Maskawa (CKM) matrix which needs additional off-diagonal terms it is necessary to consider higher order effects which are generated with different powers of the flavor scale as in Eq.~(\ref{flavon}).
Here the $U(1)_X$ quantum numbers are suitably assigned to the fields content as in Table \ref{reps}, where $p$ and $q$ are arbitrary real numbers.

In the effective theory valid below the new physics scale $\Lambda$, the quark and the lepton Yukawa couplings are functions of the SM gauge singlet scalar flavon fields $\Theta$ and $\chi$. The Yukawa matrices can be expanded in powers of the flavon fields ${\cal F}=(\Theta,\chi)$ schematically as
 \begin{eqnarray}
 Y_{ij}({\cal F})=\hat{y}_{ij}\left(\frac{{\cal F}}{\Lambda}\right)^{n}~,
 \label{flavon1}
 \end{eqnarray}
 where the $\hat{y}_{ij}$ are numerical coefficients.

We assume the following hierarchy of scales: $v_{\chi}$ and $M$ (seesaw scale) are much larger than $v_{\Phi}$ and $v_{\eta}$ (electroweak scale) and much less than $v_{\Theta}$ and $\Lambda$ (flavon scale),
 \begin{eqnarray}
v_{\Phi} \approx v_{\eta} \ll M \approx v_{\chi} \ll v_{\Theta} \ll \Lambda.
 \label{vevhier}
 \end{eqnarray}

According to this hierarchy, in the effective lagrangian below the flavor scale we keep only the leading order terms in $1/\Lambda$ and up to the linear terms in $\chi$. With the representation assignments in Table~\ref{reps}, the Lagrangian terms bilinear in the lepton and quark fields, invariant under $SU(2)\times U(1)\times A_{4}\times U(1)_X\times Z_{2}$, and to the order just mentioned are given by
\begin{align}
{\cal L}_{\rm leading} = {\cal L}^{u}_{\rm leading} + {\cal L}^{d}_{\rm leading} + {\cal L}^{\ell}_{\rm leading} + {\cal L}^{\nu}_{\rm leading} + {\rm h.c.},
\label{lagrangian}
\end{align}
where
\begin{align}
 -{\cal L}^{u}_{\rm leading} &
   =\frac{\hat{y}_{u}}{\Lambda^{10}}\,\Theta^{10}\,\bar{Q}_{L_1}\tilde{\eta}~u_{R}
  +\frac{\hat{y}_{uc}}{\Lambda^7}\,\Theta^7\,\bar{Q}_{L_1}\tilde{\eta}~c_{R}
  +\frac{\hat{y}_{ut}}{\Lambda^3}\,\Theta^3\,\bar{Q}_{L_1}\tilde{\eta}~t_{R}
  \nonumber\\
  &
  +\frac{\hat{y}_{cu}}{\Lambda^9}\,\Theta^9\,\bar{Q}_{L_2}\tilde{\eta}~u_{R}
  +\frac{\hat{y}_{c}}{\Lambda^6}\,\Theta^6\,\bar{Q}_{L_2}\tilde{\eta}~c_{R}
  +\frac{\hat{y}_{ct}}{\Lambda^2}\,\Theta^2\,\bar{Q}_{L_2}\tilde{\eta}~t_{R}
  \nonumber\\
  &
  +\frac{\hat{y}_{tu}}{\Lambda^7}\,\Theta^7\,\bar{Q}_{L_3}\tilde{\eta}~u_{R}
  +\frac{\hat{y}_{tc}}{\Lambda^4}\,\Theta^4\,\bar{Q}_{L_3}\tilde{\eta}~c_{R}
  +\hat{y}_{t}\,\bar{Q}_{L_3}\tilde{\eta}~t_{R}
  \label{U-lag}
\\~\nonumber\\
 -{\cal L}^{d}_{\rm leading} &
  =\frac{\hat{y}_{d}}{\Lambda^6}\,\Theta^6\,\bar{Q}_{L_1}(\Phi D_{R})_{{\bf 1}}
  +\frac{\hat{y}_{s}}{\Lambda^5}\,\Theta^5\,\bar{Q}_{L_2}(\Phi D_{R})_{{\bf 1}}
  +\frac{\hat{y}_{b}}{\Lambda^3}\,\Theta^3\,\bar{Q}_{L_3}(\Phi D_{R})_{{\bf 1}}
  \nonumber\\
  &
  +\frac{1}{\Lambda^6}\,\Theta^5\,\bar{Q}_{L_1} (\hat{x}_d \Phi D_{R} {\chi}^{\ast})_{{\bf 1}}
  +\frac{1}{\Lambda^5}\,\Theta^4\,\bar{Q}_{L_2}(\hat{x}_s \Phi D_{R}{\chi}^{\ast})_{{\bf 1}}
  +\frac{1}{\Lambda^3}\,\Theta^2\,\bar{Q}_{L_3}(\hat{x}_b\Phi D_{R}{\chi}^{\ast})_{{\bf 1}}
   \label{D-lag}
\\~\nonumber\\
 -{\cal L}^{\ell}_{\rm leading} &
  =\frac{\hat{y}_{e}}{\Lambda^8}\,\Theta^8\,\bar{L}_{e}\,\eta~e_{R}
  +\frac{\hat{y}_{\mu}}{\Lambda^5}\,\Theta^5\,\bar{L}_{\mu}\,\eta~\mu_{R}
  +\frac{\hat{y}_{\tau}}{\Lambda^3}\,\Theta^3\,\bar{L}_{\tau}\,\eta~\tau_{R}
\\~\nonumber\\
 -{\cal L}^{\nu}_{\rm leading} &
  =\frac{\hat{y}^{\nu}_{1}}{\Lambda}\,\Theta\,\bar{L}_{e}(\tilde{\Phi}N_{R})_{{\bf 1}}
  +\frac{\hat{y}^{\nu}_{2}}{\Lambda}\,\Theta\,\bar{L}_{\mu}(\tilde{\Phi}N_{R})_{{\bf 1}'}
  +\frac{\hat{y}^{\nu}_{3}}{\Lambda}\,\Theta\,\bar{L}_{\tau}(\tilde{\Phi}N_{R})_{{\bf 1}''}
  \nonumber\\
  &
  +\frac{1}{2}M(\overline{N^{c}_{R}}N_{R})_{{\bf 1}}
  +\frac{1}{2}\frac{\hat{y}^{\nu}_{R}}{\Lambda}\,\Theta\,\big[(\overline{N^{c}_{R}}N_{R})_{{\bf 3}_{s}} \chi\big]_{{\bf 1}}
  \nonumber\\
  &
  +\frac{1}{\Lambda}\,\bar{L}_{e}(\hat{x}^{\nu}_{1}\tilde{\Phi}N_{R}\chi^{\ast})_{{\bf 1}}
  +\frac{1}{\Lambda}\,\bar{L}_{\mu}(\hat{x}^{\nu}_{2}\tilde{\Phi}N_{R}\chi^{\ast})_{{\bf 1}'}
  +\frac{1}{\Lambda}\,\bar{L}_{\tau}(\hat{x}^{\nu}_{3}\tilde{\Phi}N_{R}\chi^{\ast})_{{\bf 1}''} .
  \label{N-lag}
 \end{align}
Here the $\hat{x}$ and $\hat{y}$'s are numerical coefficients,  the fields $\tilde{\Phi}\equiv i\tau_{2}\Phi^{\ast}$ and $\tilde{\eta}\equiv i\tau_{2}\eta^{\ast}$ are obtained with the help of the Pauli matrix $\tau_{2}$, and in Eqs.~(\ref{D-lag}) and (\ref{N-lag}) we have used the abbreviations
\begin{align}
(\hat{x} a b c)_{\bf 1} = \hat{x}^s [a(bc)_{{\bf 3}_s}]_{\bf 1} + \hat{x}^a [a(bc)_{{\bf 3}_a}]_{\bf 1} ,
\end{align}
and those obtained by replacing  ${\bf 1}$ with ${\bf 1}'$ or ${\bf 1}''$, for the $A_4$-singlet part of the product of three $A_4$-triplet fields $a$, $b$, and $c$.

In the Lagrangian (\ref{lagrangian}), each flavor of quarks has its own independent Yukawa term, with the same representation of $A_4$ but different representation of $U(1)_X$. Similarly, each flavor of charged-leptons has its own independent Yukawa terms, since the $A_4$ singlet charged-leptons $L_{e}~(e_R)$, $L_{\mu}~(\mu_R)$, and $L_{\tau}~(\tau_R)$ belong to different singlet representations ${\bf 1}$, ${\bf 1}'$, and ${\bf 1}''$ of $A_4$, respectively. Therefore, the up-type quark and charged lepton mass matrices are automatically diagonal due to the $A_{4}$-singlet nature of the up-type quark, charged lepton, and $SU(2)_{L}$ doublet Higgs field. The up-type quark Yukawa terms and the charged lepton terms involve the $A_{4}$ singlet Higgs $\eta$.
Each flavor of Dirac neutrinos also has its own independent Yukawa term, since they belong to different singlet representations ${\bf 1}$, ${\bf 1}'$, and ${\bf 1}''$ of $A_4$: the Dirac neutrino Yukawa terms involve the $A_{4}$ triplets $\Phi$ and $N_R$, which combine into the appropriate singlet representation.
Since the right-handed neutrinos having a mass scale much above the weak interaction scale are complete singlets of the SM gauge symmetry, it can possess bare SM invariant mass terms.
In addition to the bare mass term, the right-handed neutrinos have another independent Yukawa term that involve the $A_4$-triplet SM-singlet Higgs $\chi$.  The terms in ${\cal L}^{d}_{\rm leading}$ provide off-diagonal entries in the down-type quark mass matrix and to the two small mixing angles in the quark CKM matrix with the condition Eq.~(\ref{vevhier}). Notice that the $Z_{2}$ symmetry forbids terms of the form $\Theta^n \bar{Q}_{L_i} \tilde{\Phi} \chi^\ast$ in ${\cal L}^{u}_{\rm leading}$ and of the form $\Theta^n \bar{L}_\ell \Phi \chi^\ast$ in ${\cal L}^{\ell}_{\rm leading}$, i.e., enforces that $\eta$ and not $\Phi$ contribute mass terms to the up-quarks and charged leptons.

From the leading-order Lagrangian (\ref{lagrangian}) we obtain the following Yukawa and Majorana--mass terms for the quark and lepton fields in the effective Lagrangian below the flavon scale,
\begin{align}
-{\cal L}_{\rm Yuk} &= \bar{Q}_{L} \, Y_{u} \, \tilde{\eta} \, q^{u}_{R}
 + \bar{Q}_{L} \, Y^{(i)}_{d} \, \Phi_i \, q^{d}_{R}
  \nonumber\\&
 + \bar{L} \, Y_{\ell}  \, \eta \, \ell_{R}
 + \bar{L} \, Y^{(i)}_{\nu} \, \tilde\Phi_i  \, N_{R}
 \nonumber\\&
 + \frac{1}{2}  \overline{N_{R}^{\rm c}} \, M_{R} \, N_{R}
 + {\rm h.c.}.
\label{eq:effYuk}
\end{align}
Here $i=1,2,3$ and we use matrix notation with $Q_{L} = (Q_{L_1}, Q_{L_2}, Q_{L_3})^T$, $ q^{u} = (u, c, t)^T$, $q^{d} = (d, s, b)^T$, $L = (L_1, L_2, L_3)^T$, $\ell = (e,\mu,\tau)^T$, $N_{R} = (N_{R1},N_{R2},N_{R3})^T$.
The Yukawa matrices $Y$ and the neutrino Majorana mass matrix $M_R$ are
{\allowdisplaybreaks
\begin{gather}
Y_{u} = \begin{pmatrix}
 \lambda^{10}\,\hat{y}_{u} &  \lambda^7 \,\hat{y}_{uc} &  \lambda^3 \,\hat{y}_{ut}  \\
 \lambda^9 \,\hat{y}_{cu} & \lambda^6 \,\hat{y}_{c} & \lambda^2 \,\hat{y}_{ct} \\
 \lambda^7 \,\hat{y}_{tu} & \lambda^4 \,\hat{y}_{tc} & \hat{y}_{t}
\end{pmatrix},
\quad
Y_{\ell} = \begin{pmatrix}
 \lambda^8 \,\hat{y}_{e} & 0 & 0 \\
 0 & \lambda^5 \,\hat{y}_{\mu} & 0 \\
 0 & 0 & \lambda^3 \,\hat{y}_{\tau}
\end{pmatrix},\label{Yuk:Yl}
\\
Y^{(1)}_{d} =  \begin{pmatrix}
 \lambda^6 \,\hat{y}_{d} & 0 & 0 \\
 \lambda^5 \,\hat{y}_{s} & 0 & 0 \\
 \lambda^3 \,\hat{y}_{b} & 0 & 0
\end{pmatrix} ,
\quad
Y^{(2)}_{d} = \begin{pmatrix}
 0 & \lambda^6 \,\hat{y}_{d} & \epsilon \lambda^7 \,\hat{x}^{+}_{d}  \\
 0 & \lambda^5 \,\hat{y}_{s} & \epsilon \lambda^6 \,\hat{x}^{+}_{s} \\
 0 & \lambda^3 \,\hat{y}_{b} & \epsilon \lambda^4 \,\hat{x}^{+}_{b}
\end{pmatrix} ,
\quad
Y^{(3)}_{d} =  \begin{pmatrix}
 0 & \epsilon \lambda^7 \,\hat{x}^{-}_{d} & \lambda^6 \,\hat{y}_{d} \\
 0 & \epsilon \lambda^6 \,\hat{x}^{-}_{s} & \lambda^5 \,\hat{y}_{s} \\
 0 & \epsilon \lambda^4 \,\hat{x}^{-}_{b}  & \lambda^3 \,\hat{y}_{b}
\end{pmatrix} ,
\label{Yuk:Yd}
\\
Y^{(1)}_{\nu} = \begin{pmatrix}
 \lambda \,\hat{y}_{1}^{\nu} & 0 & 0 \\
 \lambda \,\hat{y}_{2}^{\nu} & 0 & 0 \\
 \lambda \,\hat{y}_{3}^{\nu} & 0 & 0
\end{pmatrix} ,
\quad
Y^{(2)}_{\nu} = \begin{pmatrix}
 0 & \lambda \hat{y}_{1}^{\nu} & \epsilon \lambda^2 \hat{x}_{1}^{\nu+} \\
 0 & \omega^2 \lambda \hat{y}_{2}^{\nu} & \omega^2 \epsilon \lambda^2 \hat{x}_{2}^{\nu+} \\
 0 & \omega \lambda \hat{y}_{3}^{\nu} & \omega \epsilon \lambda^2 \hat{x}_{3}^{\nu+}
\end{pmatrix} ,
\quad
Y^{(3)}_{\nu} = \begin{pmatrix}
 0 & \epsilon \lambda^2 \hat{x}_{1}^{\nu-} & \lambda \hat{y}_{1}^{\nu} \\
 0 & \omega \epsilon \lambda^2 \hat{x}_{2}^{\nu-} & \omega \lambda \hat{y}_{2}^{\nu} \\
 0 & \omega^2 \epsilon \lambda^2 \hat{x}_{3}^{\nu-} & \omega^2 \lambda \hat{y}_{3}^{\nu}
\end{pmatrix} ,
\\
M_{R} = \begin{pmatrix}
  M & 0 & 0 \\
  0 & M & \kappa\,M  \\
  0 & \kappa\,M & M
  \end{pmatrix} .\label{Yuk:MR}
\end{gather}
}

\noindent Here we have defined $\hat{x}^{\pm}_{f} = \hat{x}^{s}_{f} \pm \hat{x}^{a}_{f}$,  the complex parameter
\begin{align}
\kappa = \hat{y}_{R}^{\nu} \, \frac{v_\chi}{M} \, \frac{v_\Theta}{\Lambda} ,
\end{align}
and the parameters $\lambda$ (the Cabibbo angle parameter) and $\epsilon$
 \begin{eqnarray}
 \lambda = \frac{v_{\Theta}}{\Lambda}, \quad \epsilon = \frac{1}{\lambda} \, \frac{v_{\chi}}{v_{\Theta}}.
 \label{vevhier2}
 \end{eqnarray}
In the hierarchy (\ref{vevhier}), $\epsilon\approx\lambda \ll 1$.

Notice that the effective Lagrangian in Eq.~(\ref{eq:effYuk}), with complex coefficients $\hat{y}$ and $\hat{x}$ of order one, can be taken as the starting point of our model.
Inspection of Eqs.~(\ref{Yuk:Yl}--\ref{Yuk:MR}) shows that the top quark has its own renormalizable Yukawa coupling and the Majorana nuetrino has bare mass term and each Dirac neutrino has the same order of magnitude of Yukawa coupling, while other couplings are suppressed by successive powers of ${\cal F}/\Lambda$. This supplies the the strong and mild hierarchical Yukawa couplings needed to explain the charged fermion masses and the light neutrino masses, respectively.

To summarize, the flavon-fermion couplings and the expansion in inverse powers of the large scale $\Lambda$ has the following consequences.
\begin{itemize}
  \item[(i)] All Yukawa couplings $\hat{y}$ appearing in the Lagrangian~(\ref{lagrangian}) are complex numbers of order $\approx 1$. Non-renormalizable terms appear with successive powers of the flavor fields ${\cal F}$.
  \item[(ii)] The neutrino mass terms arise from the first term in Eq.~(\ref{lagrangian}), which is renormalizable, as well as the third term, which is non-renormalizable but the corresponding Yukawa couplings have the same form $\hat{y}^{\nu}_{i}\Theta/\Lambda$ $(i=1,2,3)$, thus explaining why the hierarchy of neutrino masses is mild. The charged fermion mass terms arise from the sum of the first (renormalizable) and last three terms (non-renormalizable and containing the heavy mass scale $\Lambda$), thus describing why the hierarchy of the charged fermion masses is strong.
  \item[(iii)] By integrating out the heavy flavor fields, all effective Yukawa couplings become hierarchical Yukawa couplings, and the $U(1)_X$ charge assignments make them correspond to the measured fermion mass hierarchies.
\end{itemize}

After electroweak and $A_4$ symmetry breaking, the neutral Higgs fields acquire vacuum expectation values and give masses to the fermions. The Higgs doublet $\eta$ gives masses to the up-type quarks and the charge leptons, the Higgs doublet $\Phi$ gives Dirac masses to the three SM neutrinos, and the flavon Higgs singlet $\chi$ give Majorana masses to the right-handed neutrinos. These Majorana masses are large and lead to the seesaw mechanism for neutrino masses.

\section{Mass matrices and Mixing matrices}
The SU(2) electroweak symmetry is spontaneously broken by nonzero vacuum expectation values for the Higgs fields $\eta$ and $\Phi_i$ ($i=1,2,3$). As explained in Appendix \ref{sec:minSU2}, the vacuum alignment
 \begin{align}
  \langle\eta\rangle=\frac{v_{\eta}}{\sqrt{2}},
  \quad
  \langle\Phi_i\rangle=\frac{v_{\Phi}}{\sqrt{2}}
 \end{align}
provides a minimum of the electroweak Higgs potential.
The SM VEV $v_{\rm EW}=(\sqrt{2}G_{F})^{-1/2}=246$ GeV results from the combination
\begin{align}
v_{\rm EW}=\sqrt{v^{2}_{\eta}+3v^{2}_{\Phi}}.
\end{align}
In our numerical calculations, we set
 \begin{eqnarray}
  v_{\Phi} = v_{\eta}=123\,{\rm GeV}.
 \label{vevhier1}
 \end{eqnarray}

In the following, we use the matrix notation $q^{u}=(u,c,t)$, $q^{d}=(d,s,b)$, $\ell=(e,\mu,\tau)$, $\nu=(\nu_e,\nu_\mu,\nu_\tau)$, and $N_R=(N_{R1},N_{R2},N_{R3})$. We recall that the quark and lepton fields in the lagrangian are weak interaction eigenstates, i.e., the charged-current interaction term reads
\begin{align}
-{\cal L}_{\rm c.c.} = \frac{g}{\sqrt{2}}W^{+}_{\mu} ~\overline{q^{u}_{L}}\gamma^{\mu}q^{d}_{L} +\frac{g}{\sqrt{2}}W^{-}_{\mu}\overline{\ell_{L}}\gamma^{\mu}\nu_{L}+ {\rm h.c.} ~,
\end{align}
where $g$ is the SU(2) coupling constant.

\subsection{Quark sector}
The quark mass terms can be written in matrix form as
 \begin{eqnarray}
 -{\cal L} &=& \overline{q^{u}_{L}}\mathcal{M}_{u}q^{u}_{R}+\overline{q^{d}_{L}}\mathcal{M}_{d}q^{d}_{R}
+ {\rm h.c.} ~.
 \label{lagrangianChA}
 \end{eqnarray}
Here
\begin{align}
\mathcal{M}_{u} = Y_u \langle \eta^0 \rangle ,
\quad
\mathcal{M}_{d} = \sum_{i=1}^{3} Y^{(i)}_{d} \langle \Phi^{0}_i \rangle .
\end{align}
Explicitly, for $\langle \eta^0 \rangle = v_\eta/\sqrt{2}$ and $\langle \Phi^{0}_i \rangle = v_\Phi/\sqrt{2}$,
\begin{align}
  \mathcal{M}_{u}&=\frac{v_{\eta}}{\sqrt{2}}
\begin{pmatrix}
 \lambda^{10}\,\hat{y}_{u} &  \lambda^7 \,\hat{y}_{uc} &  \lambda^3 \,\hat{y}_{ut}  \\
 \lambda^9 \,\hat{y}_{cu} & \lambda^6 \,\hat{y}_{c} & \lambda^2 \,\hat{y}_{ct} \\
 \lambda^7 \,\hat{y}_{tu} & \lambda^4 \,\hat{y}_{tc} & \hat{y}_{t}
\end{pmatrix}
 \label{eq:Mu}\\
 \intertext{and}
\mathcal{M}_{d} &=\frac{v_{\Phi}}{\sqrt{2}}\lambda^3
\begin{pmatrix} \lambda^3 \,\hat{y}_{d} & \lambda^3 \,\hat{y}_{d} & \lambda^3 \,\hat{y}_{d} \\
 \lambda^2 \,\hat{y}_{s} &  \lambda^2\,\hat{y}_{s} &  \lambda^2\,\hat{y}_{s} \\
 \hat{y}_{b} & \hat{y}_{b} &  \hat{y}_{b}\end{pmatrix}
 +
 \frac{v_{\Phi}}{\sqrt{2}}\,\epsilon\,\lambda^4 \begin{pmatrix}
0 &  \lambda^3\,\hat{x}^{-}_{d}  & \lambda^3\,\hat{x}^{+}_{d} \\
0 &  \lambda^2\,\hat{x}^{-}_{s} & \lambda^2\,\hat{x}^{+}_{s} \\
0 &  \hat{x}^{-}_{b} &  \hat{x}^{+}_{b}\end{pmatrix},
\label{eq:Md}
\end{align}
Recalling that all the hat Yukawa couplings appearing in Eqs.~(\ref{eq:Mu}) and (\ref{eq:Md}) are of order unity and arbitrary complex numbers, and the magnitude of $\epsilon$ should not be very small in order to generate the correct CKM matrix.
The mass terms in Eq.~(\ref{lagrangianChA}) indicate that, with the VEV alignments in Eqs.~(\ref{eq:VEVchitheta}) and (\ref{Alin2}), the $A_{4}$ symmetry is spontaneously and completely broken and there is no residual symmetry from $A_4$.

The up (down)-type quark mass matrix $\mathcal{M}_{f}$ with $f=u,d$ can be diagonalized in the mass basis by a biunitary transformation,
 \begin{eqnarray}
  \widehat{\mathcal{M}}_{f}=V^{f\dag}_{L}\mathcal{M}_{f}V^{f}_{R}=\diag(m_{f_1},m_{f_2},m_{f_3})\,,
 \end{eqnarray}
by the field redefinitions $q^{f}_{L}\rightarrow V^{f\dag}_{L}q^{f}_{L}$ and $q^{f}_{R}\rightarrow V^{f\dag}_{R}q^{f}_{R}$. Here, the unitary matrices $V^{f}_{L}$ and $V^{f}_{R}$ can be determined by diagonalizing the Hermitian matrices ${\cal M}_{f}{\cal M}^{\dag}_{f}$ and ${\cal M}^{\dag}_{f}{\cal M}_{f}$, respectively. (Here a general $3\times3$ diagonalizing mixing matrix is given in Eq.~(\ref{Vl}))
Especially, the left-handed up (down)-type quark mixing matrices $V^{f}_{L}$ becomes one of the matrices composing the CKM matrix such as $V_{\rm CKM}=V^{u\dag}_{L}V^{d}_{L}$ (see Eq.~(\ref{CKM}) below).

In fact, consider the both matrices in Eqs.~(\ref{eq:Mu},\ref{eq:Md}) to obtain the CKM matrix and the quark masses.
In the up-type quark sector, the left-handed up-type quark mixing matrix $V^{u}_{L}$, diagonalizing the Hermitian matrix ${\cal M}_{u}{\cal M}^{\dag}_{u}$, can be obtained by
 \begin{align}
 &V^{u\dag}_{L}{\cal M}_{u}{\cal M}^{\dag}_{u}V^{u}_{L}
=\begin{pmatrix}
m^{2}_{u} & 0 & 0 \\
 0 & m^{2}_{c} & 0 \\
 0 &  0 & m^{2}_{t} \end{pmatrix}\nonumber\\
 &\simeq \frac{v_{\eta}^2}{2}V^{u\dag}_{L}\begin{pmatrix}
 \lambda^{6}|\hat{y}_{ut}|^2+\lambda^{14}|\hat{y}_{uc}|^2 & \lambda^{5}\hat{y}_{ut}\,\hat{y}^{\ast}_{ct}+\lambda^{13}\hat{y}_{uc}\,\hat{y}^{\ast}_{c}  & \lambda^{3}\hat{y}_{ut}\,\hat{y}^{\ast}_{t}+\lambda^{11}\hat{y}_{uc}\,\hat{y}^{\ast}_{tc}  \\
 \lambda^{5}\hat{y}^{\ast}_{ut}\,\hat{y}_{ct}+\lambda^{13}\hat{y}^{\ast}_{uc}\,\hat{y}_{c}  & \lambda^{4}|\hat{y}_{ct}|^2+\lambda^{12}|\hat{y}_{c}|^2 & \lambda^{2}\hat{y}_{ct}\,\hat{y}^{\ast}_{t}+\lambda^{10}\hat{y}_{c}\,\hat{y}^{\ast}_{tc}   \\
 \lambda^{3}\hat{y}^{\ast}_{ut}\,\hat{y}_{t}+\lambda^{11}\hat{y}^{\ast}_{uc}\,\hat{y}_{tc}  & \lambda^{2}\hat{y}^{\ast}_{ct}\,\hat{y}_{t}+\lambda^{10}\hat{y}^{\ast}_{c}\,\hat{y}_{tc} & |\hat{y}_{t}|^2+\lambda^{8}|\hat{y}_{tc}|^2
 \label{MMU}
\end{pmatrix}V^{u}_{L}
 \end{align}
(Here we do not display the largest power of $\lambda$ in each entry of $\mathcal{M}_{u}\mathcal{M}^{\dag}_{u}$.)
Under the constraint of unitarity, the left-handed mixing matrix $V^u_L$ can be approximated due to the strong hierarchy in Eq.~(\ref{eq:Mu}) as
 \begin{eqnarray}
 V^{u}_{L}\simeq
{\left(\begin{array}{ccc}
 1 & 0  & \lambda^{3}\frac{|\hat{y}_{ut}|}{|\hat{y}_t|}e^{i\phi^{u}_{2}}  \\
 0 &  1 & \lambda^{2}\frac{|\hat{y}_{ct}|}{|\hat{y}_t|}e^{i\phi^{u}_{1}} \\
 - \lambda^{3}\frac{|\hat{y}_{ut}|}{|\hat{y}_t|}e^{-i\phi^{u}_{2}} &  - \lambda^{2}\frac{|\hat{y}_{ct}|}{|\hat{y}_t|}e^{-i\phi^{u}_{1}} & 1
 \end{array}\right)}Q_{u} +{\cal O}(\lambda^{5}) ~,
 \label{upMixing}
 \end{eqnarray}
where $\phi^u_{1}\approx\frac{1}{2}\arg(\hat{y}_{ct}\hat{y}^{\ast}_{t}),\phi^u_{2}\approx\frac{1}{2}\arg(\hat{y}_{ut}\hat{y}^{\ast}_{t})-\frac{1}{4}\arg(\hat{y}_{ct}\hat{y}^{\ast}_{t})$, and a diagonal phase matrix $Q_u={\rm diag}(e^{i\xi^{u}_{1}},e^{i\xi^{u}_{2}},e^{i\xi^{u}_{3}})$, which can be rotated away by the redefinition of left-handed up-type quark fields.
And the corresponding mass eigenvalues of the up-type quark are given by
\begin{eqnarray}
  m_{u}\approx\frac{v_\eta}{\sqrt{2}} |\hat{y}_{uc}| \lambda^{7},\quad m_{c}\approx\frac{v_\eta}{\sqrt{2}} \frac{|\hat{y}_{ct}|^2}{\sqrt{3}|\hat{y}_{t}|} \lambda^{4},\quad m_{t}\approx\frac{v_\eta}{\sqrt{2}} | \hat{y}_t|. \label{UPmass}
 \end{eqnarray}

Similarly, in the down-type quark sector, $\mathcal{M}_{d}$ in Eq.~(\ref{eq:Md}) generates the down-type quark masses and their corresponding mixing parameters. In order to diagonalize the matrix $\mathcal{M}_{d}$, we consider the Hermitian matrix $\mathcal{M}_{d}\mathcal{M}^{\dag}_{d}$ from which we obtain the masses and mixing matrices through diagonalization: we have, showing the leading power of $\lambda$ explicitly as derived from the behavior of the Yukawa coefficients in Eq.~(\ref{eq:Md}),
 \begin{align}
 V^{d\dag}_{L}\mathcal{M}_{d}\mathcal{M}^{\dag}_{d}V^{d}_{L} &= \frac{v^2_{\Phi}}{2}\lambda^{6}V^{d\dag}_{L}
 \begin{pmatrix}
 \hat{M}_{11} \lambda^{6} &  \hat{M}_{12} \lambda^{5} &  \hat{M}_{13} \lambda^{3} \\
 \hat{M}^{\ast}_{12} \lambda^{5} &  \hat{M}_{22} \lambda^{4} &  \hat{M}_{23} \lambda^{2} \\
 \hat{M}^{\ast}_{13} \lambda^{3} &  \hat{M}^{\ast}_{23} \lambda^{2} & \hat{M}_{33}
 \end{pmatrix}V^{d}_{L}=\begin{pmatrix} m^{2}_{d} & 0 & 0 \\
 0 & m^{2}_{s} & 0 \\
 0 &  0 & m^{2}_{b} \end{pmatrix}.
 \end{align}
 Here
  \begin{eqnarray}
  \hat{M}_{11}&=& |\hat{y}_{d}|^2+|\tilde{y}_{d}|^2+|\tilde{x}_{d}|^2 ,\qquad\qquad
  \hat{M}_{22}= |\hat{y}_{s}|^2+|\tilde{y}_{s}|^2+|\tilde{x}_{s}|^2 ,\nonumber\\
  \hat{M}_{33}&=& |\hat{y}_{b}|^2+|\tilde{y}_{b}|^2+|\tilde{x}_{b}|^2 ,\qquad\qquad
  \hat{M}_{12}= \hat{y}_{d}\hat{y}^{\ast}_{s}+\tilde{y}_{d}\tilde{y}^{\ast}_{s}+\tilde{x}_{d}\tilde{x}^{\ast}_{s},\nonumber\\
  \hat{M}_{13}&=&  \hat{y}_{d}\hat{y}^{\ast}_{b}+\tilde{y}_{d}\tilde{y}^{\ast}_{b}+\tilde{x}_{d}\tilde{x}^{\ast}_{b}, \qquad\qquad
  \hat{M}_{23}= \hat{y}_{s}\hat{y}^{\ast}_{b}+\tilde{y}_{s}\tilde{y}^{\ast}_{b}+\tilde{x}_{s}\tilde{x}^{\ast}_{b},
 \label{Mele}
 \end{eqnarray}
with $\tilde{y}_{f}=\hat{y}_{f}+\hat{x}^{-}_{f}\epsilon\lambda$ and $\tilde{x}_{f}=\hat{y}_{f}+\hat{x}^{+}_{f}\epsilon\lambda$ with $f=d,s,b$. Recalling that $\hat{x}^{\pm}_{f} = \hat{x}^{s}_{f} \pm \hat{x}^{a}_{f}$.
The mixing matrix diagonalizing the Hermitian matrix $\mathcal{M}_{d}\mathcal{M}^{\dag}_{d}$ can be obtained as
 \begin{eqnarray}
 V^{d}_{L}={\left(\begin{array}{ccc}
 1-\frac{\lambda^2}{2}\frac{|\beta|^2}{\hat{M}^2_{22}} &  \lambda\frac{|\beta|}{\hat{M}_{22}}e^{i\phi^{d}_{3}} &  \lambda^3\frac{|\hat{M}_{13}|}{\hat{M}_{33}}e^{i\phi^{d}_{2}} \\
 -\lambda\frac{|\beta|}{\hat{M}_{22}}e^{-i\phi^{d}_{3}} &  1-\frac{\lambda^2}{2}\frac{|\beta|^2}{\hat{M}^2_{22}} &  \lambda^2\frac{|\hat{M}_{23}|}{\hat{M}_{33}}e^{i\phi^{d}_{1}} \\
 \lambda^3\left(\frac{|\hat{M}_{23}|}{\hat{M}_{33}}\frac{|\beta|}{\hat{M}_{22}}e^{-i(\phi^{d}_{1}+\phi^{d}_{3})}-\frac{|\hat{M}_{13}|}{\hat{M}_{33}}e^{-i\phi^{d}_{2}}\right) &  -\lambda^2\frac{|\hat{M}_{23}|}{\hat{M}_{33}}e^{-i\phi^{d}_{1}} &  1
 \end{array}\right)}Q_{d}+{\cal O}(\lambda^{4}) ~,
 \label{downMixing}
 \end{eqnarray}
where $\phi^{d}_{1}=\frac{1}{2}\arg(\hat{M}_{23}), \phi^d_{2}\approx\frac{1}{2}\arg(M_{13})-\frac{1}{2}\phi^d_{1}$ and $\phi^{d}_3=\frac{1}{2}\arg(\beta)-\frac{1}{2}\phi^d_{2}$ with $\beta\approx\hat{M}_{12}e^{i\phi^d_{1}}-\frac{\hat{M}_{13}|\hat{M}_{23}|}{\hat{M}_{33}}e^{-i\phi^d_{1}}$.
Here the diagonal phase matrix can be rotated away by the redefinition of left-handed down-type quark fields. As a result, the corresponding mass eigenvalues of down-type quarks are given as
\begin{eqnarray}
  m_{d}\approx\frac{v_\Phi}{\sqrt{2}} \lambda^{6} \left(\hat{M}_{11}-\frac{|\hat{M}_{13}|^2}{\hat{M}_{33}}-\frac{|\beta|^2}{\hat{M}_{22}}\right)^{1/2} ,\quad m_{s}\approx\frac{v_\Phi}{\sqrt{2}}\lambda^{5}\sqrt{\hat{M}_{22}} ,\quad m_{b}\approx\frac{v_\Phi}{\sqrt{2}}\lambda^3 \sqrt{\hat{M}_{33}}\,, \label{Downmass}
 \end{eqnarray}
where $\hat{M}_{ij}$ are numerical coefficients of order $\approx 1$ given in Eq.~(\ref{Mele}).
This provides the mass hierarchy
 \begin{align}
 m_d \approx \lambda^6 \, m_t,
 \quad
 m_s \approx \lambda^5 \, m_t,
 \quad
 m_b \approx \lambda^3 \, m_t.
\label{DOWNmass}
 \end{align}

From the charged current term in Eq.~(\ref{lagrangianChA}) we obtain the CKM matrix by combining Eq.~(\ref{upMixing}) and Eq.~(\ref{downMixing})
 \begin{eqnarray}
  V_{\rm CKM}&=&V^{u\dag}_{L}V^{d}_{L}\nonumber\\
  &=&{\left(\begin{array}{ccc}
 1-\frac{\lambda^2}{2}\frac{|\beta|^2}{\hat{M}^2_{22}} &  \lambda\frac{|\beta|}{\hat{M}_{22}}e^{i\phi^{d}_{3}} &  \lambda^3\,B\,e^{i\phi_{B}} \\
 -\lambda\frac{|\beta|}{\hat{M}_{22}}e^{-i\phi^{d}_{3}} &  1-\frac{\lambda^2}{2}\frac{|\beta|^2}{\hat{M}^2_{22}} &  \lambda^2\,A\,e^{i\phi_{A}} \\
 \lambda^3(A\frac{|\beta|}{\hat{M}_{22}}e^{-i(\phi_{A}+\phi^{d}_{3})}-B\,e^{-i\phi_{B}}) &  -\lambda^2\,A\,e^{-i\phi_{A}} &  1
 \end{array}\right)}+{\cal O}(\lambda^{4})\,.
  \label{CKM}
 \end{eqnarray}
Here $A$ and $B$ are real numbers
 \begin{eqnarray}
  A\,e^{i\phi_A}\equiv\frac{|\hat{M}_{23}|}{\hat{M}_{33}}e^{i\phi^{d}_{1}}-\frac{|\hat{y}_{ct}|}{|\hat{y}_{t}|}e^{i\phi^{u}_{1}}\,,\qquad  B\,e^{i\phi_B}\equiv\frac{|\hat{M}_{13}|}{\hat{M}_{33}}e^{i\phi^{d}_{2}}-\frac{|\hat{y}_{ut}|}{|\hat{y}_{t}|}e^{i\phi^{u}_{2}}\,.
 \end{eqnarray}
It shows directly that can generate a large Cabbibo angle
$\theta_{C}\sim\lambda$ and the two small mixing angles $\theta^{q}_{13}\sim\lambda^3$ and $\theta^{q}_{23}\sim\lambda^2$.
From Eq.~(\ref{CKM}), after the field redefinitions $s_{L}\rightarrow s_{L}e^{-i\phi^{d}_{3}}$, $b_{L}\rightarrow b_{L}e^{-i(\phi_{A}+\phi^{d}_{3})}$, $c_{L}\rightarrow c_{L}e^{-i\phi^{d}_{3}}$ and $t_{L}\rightarrow t_{L}e^{-i(\phi_{A}+\phi^{d}_{3})}$, if one set
 \begin{eqnarray}
  B\,e^{-i(\phi_{A}+\phi^{d}_{3}-\phi_{B})}=A(\rho-i\eta),\qquad|\beta|^2\approx\hat{M}^{2}_{22}
 \end{eqnarray}
then one can obtain the CKM matrix in the Wolfenstein parametrization~\cite{Wolfenstein:1983yz} given by
 \begin{eqnarray}
  V_{\rm CKM}
  ={\left(\begin{array}{ccc}
 1-\frac{\lambda^2}{2} &  \lambda &  \lambda^3\,A(\rho-i\eta) \\
 -\lambda &  1-\frac{\lambda^2}{2} &  A\lambda^2 \\
 A\lambda^3(1-\rho-i\eta) &  -A\lambda^2 &  1
 \end{array}\right)}+{\cal O}(\lambda^{4})\,.
  \label{CKM1}
 \end{eqnarray}
As reported in Ref.~\cite{ckmfitter} the best-fit values of the parameters $\lambda$, $A$, $\bar{\rho}$,
$\bar{\eta}$ with $1\sigma$ errors are
 \begin{eqnarray}
  \lambda &=& \sin\theta_{C}=0.22457^{+0.00200}_{-0.00027}\,,\qquad\, A=0.823^{+0.025}_{-0.049}~, \nonumber\\
  \bar{\rho} &=&0.129^{+0.075}_{-0.027}\,,\qquad\qquad\qquad\qquad\bar{\eta}=0.348^{+0.037}_{-0.044}~,
 \label{CKMpara}
 \end{eqnarray}
where $\bar{\rho}=\rho(1-\lambda^{2}/2)$ and $\bar{\eta}=\eta(1-\lambda^{2}/2)$.

\subsection{Lepton sector}

The lepton mass terms can be written in (block) matrix form as
 \begin{eqnarray}
 -{\cal L}_{m} &=& \frac{1}{2}\overline{N^{c}_{R}}M_{R}N_{R}
 +\overline{\nu_{L}}m_{D}N_{R}+\overline{\ell_{L}}\mathcal{M}_{\ell}\ell_{R}+\text{h.c.}~ \\
 &=& \frac{1}{2} \begin{pmatrix} \overline{\nu_L} & \overline{N^{c}_R} \end{pmatrix} \begin{pmatrix} 0 & m_D \\ m_D^T & M_R \end{pmatrix} \begin{pmatrix} \nu^{c}_L \\ N_R \end{pmatrix} + \overline{\ell_{L}}\mathcal{M}_{\ell}\ell_{R}+\text{h.c.},
 \label{lagrangianA}
 \end{eqnarray}
where
\begin{align}
\mathcal{M}_{\ell} = Y_{\ell} \langle \eta^0 \rangle ,
\quad
m_{D} = \sum_{i=1}^{3} Y^{(i)}_{\nu} \langle \Phi^{0}_i \rangle .
\end{align}
Explicitly, for $\langle \eta^0 \rangle = v_\eta/\sqrt{2}$ and $\langle \Phi^{0}_i \rangle = v_\Phi/\sqrt{2}$,
\begin{align}
\mathcal{M}_{\ell} & =\frac{v_{\eta}}{\sqrt{2}} \, \lambda^3
 \begin{pmatrix}
 \hat{y}_{e}\lambda^5 & 0 & 0 \\
 0 & \hat{y}_{\mu}\lambda^2 & 0 \\
 0 & 0 & \hat{y}_{\tau}
 \end{pmatrix},
\\
m_{D}& = \frac{v_\Phi}{\sqrt{2}}\,\lambda \begin{pmatrix}
\hat{y}^{\nu}_1 & 0 & 0 \\
0 & \hat{y}^{\nu}_2 & 0 \\
0 & 0 & \hat{y}^{\nu}_3
\end{pmatrix}
\,
\begin{pmatrix}
1 & 1 & 1 \\
1 & \omega & \omega^2 \\
1 & \omega^2 & \omega
\end{pmatrix}
+
\frac{v_\Phi}{\sqrt{2}} \, \epsilon\,\lambda^2
\begin{pmatrix}
0 & \hat{x}^{\nu-}_1 & \hat{x}^{\nu+}_1\\
0 & \hat{x}^{\nu-}_2 \omega & \hat{x}^{\nu+}_2 \omega^2 \\
0 & \hat{x}^{\nu-}_3 \omega^2 & \hat{x}^{\nu+}_3 \omega
\end{pmatrix}.
\label{eq:mD}
\end{align}

In the limit of large $M$ (seesaw mechanism),  and focusing on the mass matrix of the light neutrinos $\mathcal{M}_{\nu}$ only,
\begin{align}
-{\cal L}_{m} &= \frac{1}{2} \overline{\nu_L} \mathcal{M}_{\nu} \nu^{c}_L  + \overline{\ell_{L}}\mathcal{M}_{\ell}\ell_{R}+\text{h.c.}+\text{terms in $N_R$}
\end{align}
with
 \begin{align}
 \mathcal{M}_{\nu} & = -m_D \, M_R^{-1} \, m^T_D
\\
 & =m_0\,e^{i\pi}\begin{pmatrix}
 1+2F & (1-F)\,y_{2} & (1-F)\,y_{3} \\
 (1-F)\,y_{2} & (1+\frac{F+3G}{2})\,y^{2}_{2} & (1+\frac{F-3G}{2})\,y_{2}y_{3}  \\
 (1-F)\,y_{3} & (1+\frac{F-3G}{2})\,y_{2}y_{3} & (1+\frac{F+3G}{2})\,y^2_{3}
 \end{pmatrix}
 \nonumber \\
 & + \epsilon \lambda\,m_0\,e^{i\pi} \begin{pmatrix}
 \delta_{11} & \delta_{12} & \delta_{13} \\
 \delta_{12} & \delta_{22} & \delta_{23} \\
 \delta_{13} & \delta_{23} & \delta_{33}
 \end{pmatrix}
 + \epsilon^2 \lambda^2\,m_0\,e^{i\pi} \begin{pmatrix}
 \gamma_{11} & \gamma_{12} & \gamma_{13} \\
 \gamma_{12} & \gamma_{22} & \gamma_{23} \\
 \gamma_{13} & \gamma_{23} & \gamma_{33}
 \end{pmatrix}\,,
 \label{meff}
 \end{align}
 where the parameters at the leading order are defined as
 \begin{eqnarray}
  m_{0}=\frac{v^{2}_{\Phi}|y^{\nu}_{1}|^2}{2M}\,,\quad F=\frac{1}{\kappa+1}\,,\quad G=\frac{1}{\kappa-1}\,,\quad y_{2}\equiv \frac{\hat{y}^{\nu}_{2}}{\hat{y}^{\nu}_{1}}\,,\quad  y_{3}\equiv \frac{\hat{y}^{\nu}_{3}}{\hat{y}^{\nu}_{1}}\,,
  \label{kappa}
 \end{eqnarray}
here $\kappa\equiv\tilde{\kappa}\,e^{i\phi}$ with $\tilde{\kappa}\equiv\lambda\,|\hat{y}^\nu_R|\frac{v_{\chi}}{M}$ and $\phi\equiv\arg\left(\hat{y}^{\nu}_{R}\right)$,
 and the other parameters are defined in Eqs.~(\ref{C1}) and (\ref{C2}).
We have used
\begin{align}
M_R^{-1} = \frac{1}{M(1-\kappa^2)}
\begin{pmatrix}
1-\kappa^2 & 0 & 0 \\
0 & 1 & -\kappa \\
0 & -\kappa & 1
\end{pmatrix}
= \frac{1}{M}
\begin{pmatrix}
1 & 0 & 0 \\
0 & \frac{F-G}{2} & \frac{F+G}{2} \\
0 & \frac{F+G}{2} & \frac{F-G}{2}
\end{pmatrix}\,.
\end{align}
Note here that, taking $0.6\lesssim v_{\chi}/M\lesssim3$ due to $v_{\chi}\sim M$ in Eq.~(\ref{vevhier}), $\lambda=0.225$ and $0.3\lesssim|\hat{y}^{\nu}_{R}|\lesssim3$, the value of $\tilde{\kappa}$ lies in the range $0.04\lesssim\tilde{\kappa}\lesssim2.0$. And it is expected that the masses and mixing angles are not crucially corrected by the next leading order terms due to both $\epsilon\lambda\approx\lambda^2$ and the parameters $\delta_{ij}, \gamma_{ij}$ in Eq.~(\ref{meff}) being of order unity. Since the corrections can be kept few percent level, deviations from the leading order corrections are obtained for all measurable quantities at approximately the same level. So, in what follow we take only the leading contribution.
Notice that the mass scale $m_0$ incorporates the seesaw mechanism. Notice also that once $m_{0}$ is matched to the experimental data, the value of
$y^{\nu}_{1} = \hat{y}^{\nu}_{1}\lambda$ depends sensitively on the scale $M$. For $m_{0}\simeq0.03$ eV, if the value of $\hat{y}^{\nu}_{1}$ is of order one, i.e.\ $0.3\lesssim|\hat{y}^{\nu}_{1}|\lesssim3$, the seesaw (leptogenesis) scale $M$ is in the range $2.3\times10^{12}~{\rm GeV}\lesssim\,M\lesssim2.3\times10^{14}~{\rm GeV}$.

We perform basis rotations from weak  to mass eigenstates in the leptonic sector,
\begin{eqnarray}
 \widehat{\ell}_{L} = P^{\ast}_{\ell}\ell_{L}~,\quad \widehat{\ell}_{R}= P^{\ast}_{\ell}\ell_{R}~,\quad \widehat{\nu}_{L} = U^{\dag}_{\nu}P^{\ast}_{\nu}\nu_{L}~,
 \label{rebasing}
\end{eqnarray}
where $P_{\ell}$ and $P_{\nu}$ are phase matrices and $U_\nu$ is a unitary matrix chosen so as the matrices
\begin{align}
\widehat{\mathcal{M}}_{\ell} &=P^{\ast}_{\ell}\mathcal{M}_{\ell}P_{\ell}=\diag(m_{e}, m_{\mu}, m_{\tau}) \\
\widehat{\mathcal{M}}_{\nu} &= U^{\dag}_{\nu} P_\nu^* \mathcal{M}_\nu P_\nu^* U^*_{\nu} =\diag(m_{1}, m_{2}, m_{3})
\label{diagNu}
\end{align}
are real and positive diagonal. Here $m_i$ $(i = 1,2,3)$ are the light neutrino masses.
Then from the charged current term in Eq.~(\ref{lagrangianA}) we obtain the lepton mixing matrix $U_{\rm PMNS}$ as
\begin{align}
U_{\rm PMNS}=P^{\ast}_{\ell}P_{\nu}U_{\nu}.
 \label{PMNS}
\end{align}
It is important to notice that the phase matrix $P_{\nu}$ can be rotated away by choosing the matrix $P_{\ell}=P_\nu$, i.e., by an appropriate redefinition of the left-handed charged lepton fields, which is always possible. This is an important point because the phase matrix $P_{\nu}$ accompanies the Dirac-neutrino mass matrix $m_D$, and here for simplicity we take only the leading neutrino Yukawa matrix $Y_{\nu}$ in Eq.~(\ref{eq:mD}). This means that complex phases in $Y_{\nu}$ can always be rotated away by appropriately choosing the phases of left-handed charged lepton fields. Hence without loss of generality the eigenvalues $y^{\nu}_1$, $y^{\nu}_2$, and $y^{\nu}_3$ of $Y_{\nu}$ can be real and positive.
The matrix $U_{\rm PMNS}=U_{\nu}$ can be written in terms of three mixing angles and three CP-odd phases (one for the Dirac neutrinos and two for the Majorana neutrinos) as \cite{PDG}
\begin{eqnarray}
  U_{\rm PMNS}
  &=&{\left(\begin{array}{ccc}
   c_{13}c_{12} & c_{13}s_{12} & s_{13}e^{-i\delta_{CP}} \\
   -c_{23}s_{12}-s_{23}c_{12}s_{13}e^{i\delta_{CP}} & c_{23}c_{12}-s_{23}s_{12}s_{13}e^{i\delta_{CP}} & s_{23}c_{13}  \\
   s_{23}s_{12}-c_{23}c_{12}s_{13}e^{i\delta_{CP}} & -s_{23}c_{12}-c_{23}s_{12}s_{13}e^{i\delta_{CP}} & c_{23}c_{13}
   \end{array}\right)}Q_{\nu}~,
 \label{rebasing1}
\end{eqnarray}
where $Q_{\nu}=\diag(e^{-i\varphi_{1}/2},e^{-i\varphi_{2}/2},1)$, and $s_{ij}\equiv \sin\theta_{ij}$ and $c_{ij}\equiv \cos\theta_{ij}$.
The mass matrix $\mathcal{M}_{\nu}$ is diagonalized by the PMNS mixing matrix $U_{\rm PMNS}$ as described above,
 \begin{eqnarray}
  \mathcal{M}_{\nu} &=& U_{\rm PMNS} ~\diag(m_{1},m_{2},m_{3})~ U^{T}_{\rm PMNS} .
 \end{eqnarray}
As is well-known, because of the observed hierarchy $|\Delta m^{2}_{\rm Atm}|\equiv |m^{2}_{3}-m^{2}_{1}|\gg\Delta m^{2}_{\rm Sol}\equiv m^{2}_{2}-m^{2}_{1}>0$, and the requirement of a Mikheyev-Smirnov-Wolfenstein resonance for solar neutrinos, there are two possible neutrino mass spectra: (i) the normal mass ordering (NO) $m_{1}<m_{2}<m_{3}$, and (ii) the inverted mass ordering (IO) $m_{3}<m_{1}<m_{2}$.
In the limit $y^\nu_2=y^\nu_3$ ($y_{2}\rightarrow1$), the mass matrix in Eq.~(\ref{meff}) acquires a $\mu$--$\tau$ symmetry that leads to $\theta_{13}=0$ and $\theta_{23}=-\pi/4$. Moreover, in the limit  $y^\nu_1=y^\nu_2=y^\nu_3$ ($y_{2}, y_{3}\rightarrow1$)~\footnote{In this limit there exists a neutrino  mass sum-rule~\cite{sumrule}.}, the  mass matrix~(\ref{meff}) gives the TBM~\cite{TBM} angles and their corresponding mass eigenvalues
 \begin{eqnarray}
 &&\sin^{2}\theta_{12}=\frac{1}{3},~\qquad\sin^{2}\theta_{23}=\frac{1}{2},~\qquad\sin\theta_{13}=0~,\\
  \label{tibi1}
 &&m_{1}= 3m_{0}|F|~,\qquad m_{2}=3m_{0}~,\qquad m_{3}= 3m_{0}|G|~.
 \label{TBM1}
 \end{eqnarray}
 These mass eigenvalues are disconnected from the mixing angles. However, recent neutrino data, {\it i.e.} $\theta_{13}\neq0$, require deviations of $y_{2,3}$ from unity, leading to a possibility to search for CP violation in neutrino oscillation experiments.
These deviations generate relations between mixing angles and mass eigenvalues.
Therefore Eq.~(\ref{meff}) directly indicates that there could be deviations from the exact TBM if the Dirac neutrino Yukawa couplings do not have the same magnitude.

Acquiring VEV $\langle\Theta\rangle$ as in Eq.~(\ref{vevhier}), the field-dependent Yukawa couplings of the charged leptons give rise to the mass hierarchy in the charged lepton masses. From Eq.~(\ref{eq:mD}),
 \begin{eqnarray}
  \frac{m_{e}}{m_t} \approx \frac{|\hat{y}_e|}{|\hat{y}_t|} \, \lambda^8 ,
  \quad
  \frac{m_{\mu}}{m_t} \approx \frac{|\hat{y}_\mu|}{|\hat{y}_t|} \, \lambda^5 ,
  \quad
  \frac{m_{\tau}}{m_t} \approx \frac{|\hat{y}_\tau|}{|\hat{y}_t|} \, \lambda^3,
\label{Lmass}
\end{eqnarray}
with the $|\hat{y}| \approx 1$.
On the other hand, since the Yukawa couplings of the Dirac neutrinos are not a function of the flavon fields, the mild hierarchy of the light neutrino masses is naturally guaranteed with $|y^{\nu}_{1} |\approx|y^{\nu}_{2} |\approx|y^{\nu}_{3} |\approx {\cal O}(0.1)$. From Eq.~(\ref{TBM1}) we obtain
 \begin{eqnarray}
  m^{\nu}_{1}\approx m_{0},\qquad\quad~ m^{\nu}_{2}\approx m_{0},\qquad\quad m^{\nu}_{3}\approx m_{0}\,.
  \label{Numass}
 \end{eqnarray}
Note here that the above equation does not mean that the light neutrino mass spectrum is quasi-degenerate.
In the following section, we investigate this spectrum in more detail by using a numerical analysis.

We conclude this section by summarizing the hierarchical pattern of quark and lepton masses that we obtain in our model, which reproduces the observed quark and lepton mass hierarchy. To within some numerical coefficients of order one,
\begin{gather}
m_\nu \ll m_e,
\quad
m_e \approx \lambda\,m_u,
\quad
m_u \approx \lambda\,m_d,
\quad
m_d \approx \lambda\,m_s,
\\
m_s \approx m_\mu \approx \lambda\,m_c,
\quad
m_c \approx \lambda\,m_b ,
\quad
m_b \approx m_\tau \approx \lambda^3\,m_t.
\end{gather}
Alternatively,
\begin{gather}
m_e : m_\mu : m_\tau \simeq \lambda^5 : \lambda^2 : 1 , \quad
m_u : m_c : m_t \simeq \lambda^7 : \lambda^4 : 1, \;\;
\\
m_d : m_s : m_b \simeq \lambda^3 : \lambda^2 : 1,\quad
m_b/m_t \simeq \lambda^3, \quad m_\tau/m_b \simeq 1.
\end{gather}
These relations differ from those obtained in GUT SU(5) \cite{Plentinger:2007px}, and in comparison provide a better accommodation of the $m_e/m_\mu$ ratio.
This reproduces the pattern of quark and lepton masses for $\lambda \approx 0.225$.

\section{Leptonic CP violation, $0\nu\beta\beta$-decay and Leptogenesis}
In this section we investigate the observables that can be tested in the current and the next generation of experiments, and study how our model can provide a viable baryon asymmetry in the universe through leptogenesis. In detail, we consider (i) the deviations of the atmospheric mixing angle $\theta_{23}$ from its maximal value of $45^\circ$, (ii) the generation of the low energy CP-violation phase $\delta_{CP}$ (or the Jarlskog invariant $J_{CP}$) in both normal and inverted neutrino mass orderings, and (iii) the rate of neutrinoless double beta ($0\nu\beta\beta$) decay via the effective mass $|m_{ee}|=\left|\sum_{i}U^{2}_{ei}m_{i}\right|$, which is a probe of lepton number violation at low energy. Since an observation of $0\nu\beta\beta$-decay and a sufficiently accurate measurement of its half-life can provide information on lepton number violation, the Majorana vs.\ Dirac nature of neutrinos, and the neutrino mass scale and hierarchy, we show that our model is experimentally testable in the near future.

We perform a numerical analysis using the linear algebra tools that are contained in the renormalization-group evolution program of Ref.~\cite{Antusch:2005gp}.

The Daya Bay~\cite{An:2012eh} and RENO~\cite{Ahn:2012nd} experiments have
accomplished the measurement of all three neutrino mixing angles
$\theta_{12}$, $\theta_{23}$, and $\theta_{13}$, associated with three kinds of neutrino oscillation experiments.
Global fit values and $3\sigma$ intervals for the neutrino mixing angles and the neutrino mass-squared differences~\cite{GonzalezGarcia:2012sz}
are listed in Table~\ref{exp}, where $\Delta m^{2}_{\rm Sol}\equiv m^{2}_{2}-m^{2}_{1}$, $\Delta m^{2}_{\rm Atm}\equiv m^{2}_{3}-m^{2}_{1}$ for the normal mass ordering (NO), and  $\Delta m^{2}_{\rm Atm}\equiv |m^{2}_{2}-m^{2}_{3}|$ for the inverted mass ordering (IO).

\begin{table}[t]
\begin{widetext}
\begin{center}
\caption{\label{exp} Global fit of three-flavor oscillation parameters (best-fit values and $3\sigma$ intervals)~\cite{GonzalezGarcia:2012sz}. NO = normal neutrino mass ordering; IO = inverted neutrino mass ordering. The $\oplus$ indicates the presence of two local minima in the global fit.}
\begin{ruledtabular}
\begin{tabular}{ccccccccccc} &$\theta_{13}[^{\circ}]$&$\delta_{CP}[^{\circ}]$&$\theta_{12}[^{\circ}]$&$\theta_{23}[^{\circ}]$&$\Delta m^{2}_{\rm Sol}[10^{-5}{\rm eV}^{2}]$& ~$\Delta m^{2}_{\rm Atm}[10^{-3}{\rm eV}^{2}]$~~ \\
\hline
$\!\!\begin{array}{c} \text{best-fit} \\[-2ex] \text{value} \end{array}$ & $8.71$ & $265$ & $33.57$ & $41.9\oplus50.0$ & $7.45$
 & $\begin{array}{ll} 2.417\,\text{(NO)}\\[-1ex] 2.410\,\text{(IO)} \end{array}$ \\
\hline
$\!\!\begin{array}{c} 3\sigma \\[-2ex] \text{interval} \end{array}$  & ~$[7.50,9.78]$~ & ~$[0,360]$~ & ~$[31.38,36.01]$~ & ~$[37.2,54.5]$
 & $\!\![6.98,8.05]$ & $\!\!\!\!\begin{array}{ll} [2.247,2.623]\,\text{(NO)} \\[-1ex] [2.226,2.602]\,\text{(IO)}\end{array}$ \\
\end{tabular}
\end{ruledtabular}
\end{center}
\end{widetext}
\end{table}

The mass matrices $m_{D}$ and $M_{R}$ in Eq.~(\ref{meff}) contain seven parameters: $y^{\nu}_{1},M,v_{\Phi},y_{2},y_{3},\tilde{\kappa},\phi$. The first three ($y^{\nu}_{1}$, $M,$ and $v_{\Phi}$) lead to the overall neutrino scale parameter $m_{0}$. The last four ($y_2,y_3,\tilde{\kappa},\phi$) give rise to the deviations from TBM as well as the CP phases and corrections to the mass eigenvalues (see Eq.~(\ref{TBM1})).
Since the neutrino masses are sensitive to the combination $m_{0}=v^{2}_{\Phi}|y^{\nu}_{1}|^2/(2M)$, all choices of $M$ and $v_\Phi\,y^{\nu}_{1}$ with the same $v_\Phi^2|y^{\nu}_{1}|^2/M$ give identical results for the neutrino masses and mixings. Due to the magnitude of the Yukawa couplings ($|y^{\nu}_{i}|\approx0.1$), our model seesaw scale (leptogenesis scale) can be roughly determined as $M \approx 10^{12-14}$ GeV.

\begin{figure}[h]
\begin{minipage}[h]{7.5cm}
\epsfig{figure=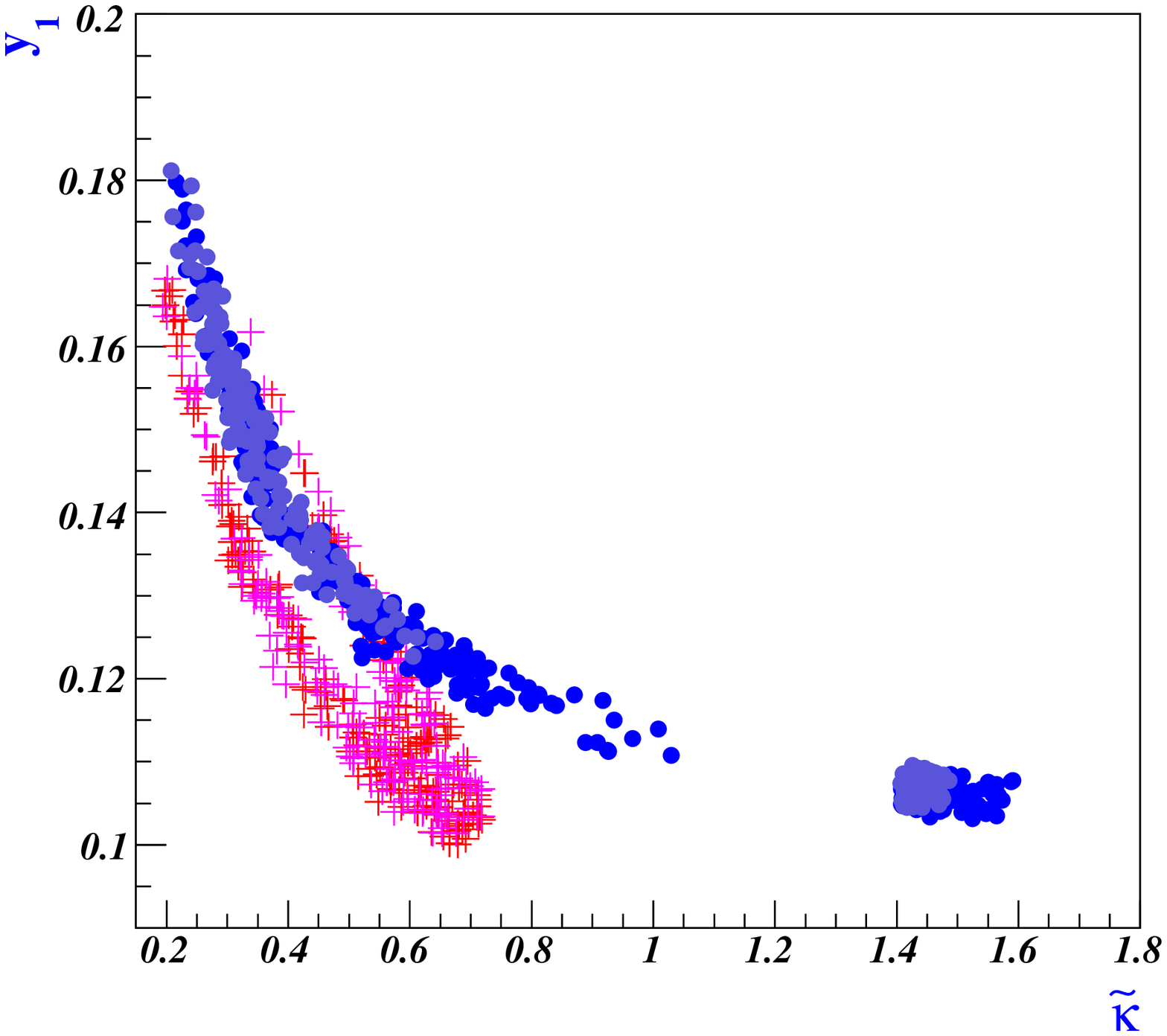,width=7.5cm,angle=0}
\end{minipage}\\
\begin{minipage}[h]{7.5cm}
\epsfig{figure=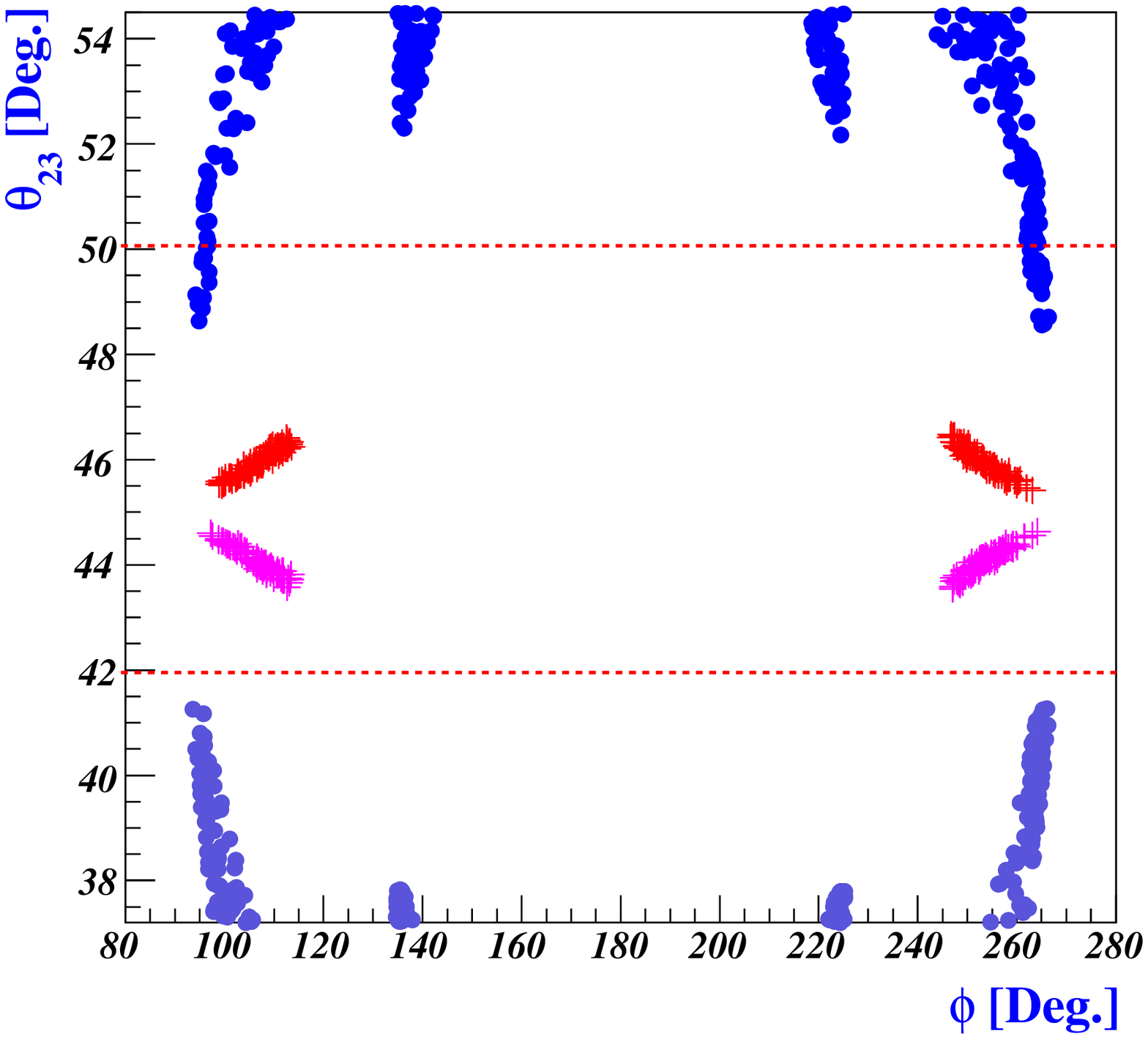,width=7.5cm,angle=0}
\end{minipage}
\begin{minipage}[h]{7.5cm}
\epsfig{figure=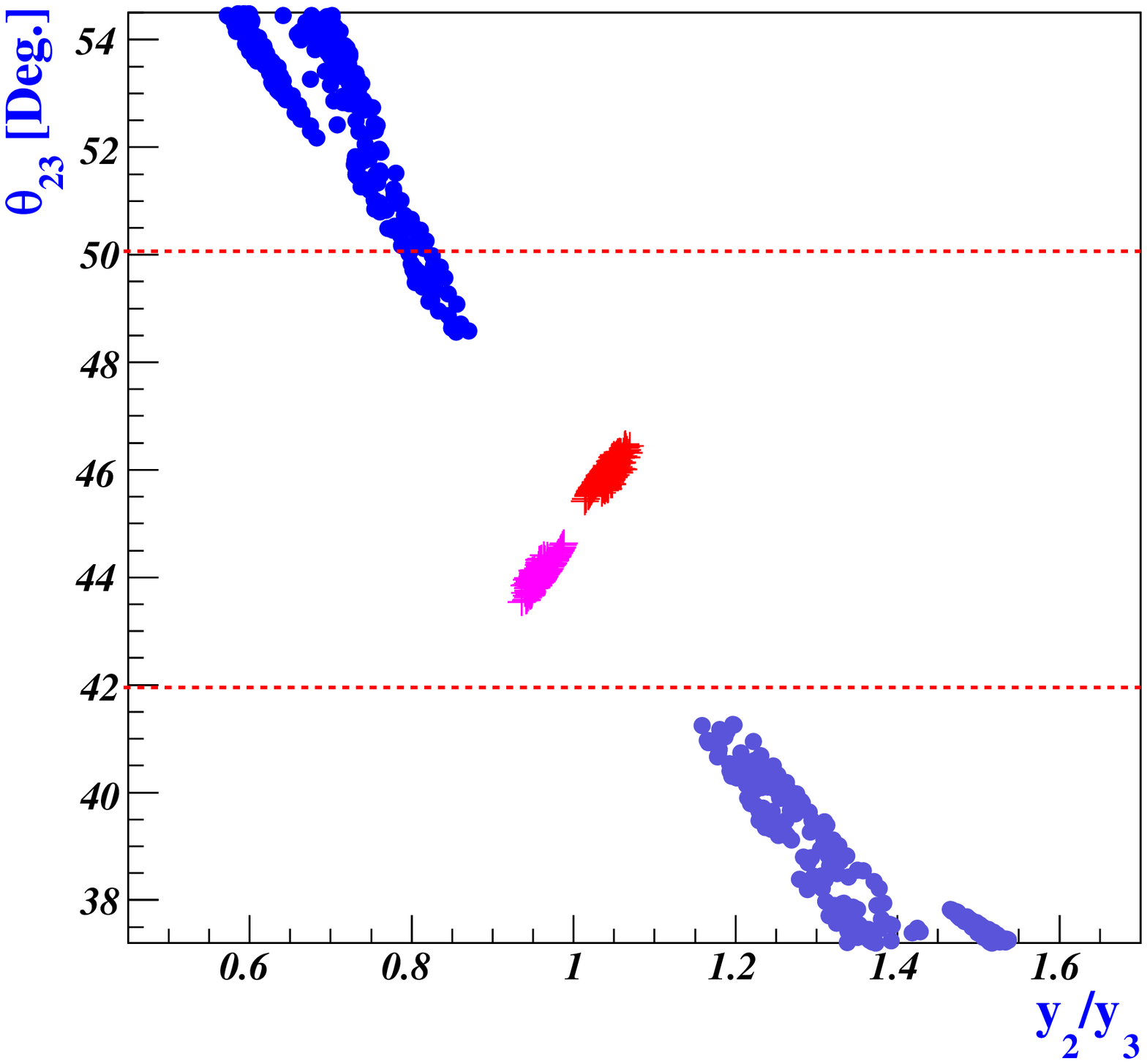,width=7.5cm,angle=0}
\end{minipage}
\caption{\label{FigA1} Scatter plots showing the location of points in parameter space lying within the $3\sigma$ experimental bounds of Table~\ref{exp}. The upper panel shows the correlation between the input parameters $\tilde{\kappa}(\equiv\lambda|y^{\nu}_{R}|v_{\chi}/M)$ and $y_{1}\equiv y^{\nu}_{1}$. The lower panels plot the dependence of the atmospheric mixing angle $\theta_{23}$ on the input parameters $\phi$ (left plot) and $y_{2}/y_{3}$ (right plot). The horizontal dotted lines show the best-fit values (two local minima). The red crosses and the blue dots correspond to normal mass ordering (NO) and inverted  mass ordering (IO), respectively.}
\end{figure}

In our numerical examples, we take $M=5\times10^{12}$ GeV and $v_{\eta}=v_{\Phi}=123$ GeV, for simplicity, as inputs.
Then the effective neutrino mass matrix in Eq.~(\ref{meff}) contains only the five parameters $m_{0},y_{2},y_{3},\tilde{\kappa},\phi$, which can be determined from five experimental results (three mixing angles $\theta_{12}$, $\theta_{13}$, and $\theta_{23}$, and two mass squared differences $\Delta m^{2}_{\rm Sol}$ and $\Delta m^{2}_{\rm Atm}$). The values of the CP-violating phases $\delta_{CP}$ and $\varphi_{1,2}$ follow after the model parameters are obtained from the experimentally measured quantities.

For given values of $M,v_\eta,v_\Phi$ we obtain the following allowed regions of the unknown model parameters within the $3\sigma$ experimental bounds in Table~\ref{exp}: for the normal mass ordering (NO),
 \begin{align}
  &\tilde{\kappa} \in [0.19,0.72] ,
  && y_{2} \in [1.0,1.25],
  \hspace{6em} & y_{3} \in [1.0,1.25],
  \nonumber\\
  & m_{0}/(10^{-2}{\rm eV}) \in [1.5,4.3],
  && \phi \in [97^{\circ},114^{\circ}]\cup[246^{\circ},265^{\circ}] ;
  \label{input1}
 \end{align}
 for the inverted mass ordering (IO),
 \begin{align}
  &\tilde{\kappa} \in [0.20,1.60] ,
  && y_{2} \in [0.74,1.25],
  \hspace{6em} y_{3} \in [0.80,1.31],
  \nonumber\\
  & m_{0}/(10^{-2}{\rm eV}) \in [1.6,4.9],
  && \phi \in [93^{\circ},113^{\circ}]\cup[134^{\circ},143^{\circ}]\cup[218^{\circ},226^{\circ}]\cup[241^{\circ},267^{\circ}] .
  \label{input2}
 \end{align}
Notice that the Dirac neutrino Yukawa couplings $y^{\nu}_{1,2,3}\approx 0.1$, and the numerical values of $\tilde{\kappa}$ lie in the range discussed below Eq.~(\ref{kappa}).

Random points in parameter space falling within the $3\sigma$ experimental bounds of Table~\ref{exp} are used to generate scatter plots showing correlations in parameter space and predictions for the observables quantities. In Fig.~\ref{FigA1} the upper panel shows the correlation between the input parameters $\tilde{\kappa}$ and $y^{\nu}_{1}$, while the lower panels plot the atmospheric mixing angle $\theta_{23}$ vs.\ the input parameters $\phi$ (left plot) and $y_{2}/y_{3}$ (right plot). Red crosses correspond to the normal mass ordering (NO) and blue dots to the inverted mass ordering (IO).

\subsection{Neutrinoless double beta ($0\nu\beta\beta$) decay}
If neutrinos are Majorana particles, an important low-energy observable is $0\nu\beta\beta$-decay, which effectively measures the absolute value of the $ee$-component of the effective neutrino mass matrix $\mathcal{M}_{\nu}$ in Eq.~(\ref{meff}) in the basis where the charged lepton mass matrix is real and diagonal:
 \begin{eqnarray}
 |m_{ee}|=\Big|\sum_{i}U^{2}_{ei}m_{i}\Big|=\Big|m_{1}c^{2}_{12}c^{2}_{13}+m_{2}s^{2}_{12}c^{2}_{13}e^{i(\varphi_{1}-\varphi_{2})}+m_{3}s^{2}_{13}e^{i(\varphi_{1}-2\delta_{CP})}\Big|~.
 \label{mee}
 \end{eqnarray}
Since the $0\nu\beta\beta$-decay is a probe of lepton number violation at low energy, its measurement could be the strongest evidence for lepton number violation at high energy. In other words, the discovery of $0\nu\beta\beta$-decay may suggest the Majorana character of the neutrinos and thus the existence of heavy Majorana neutrinos (via the seesaw mechanism), which are a crucial ingredient for leptogenesis.

Current $0\nu\beta\beta$-decay experimental upper limits and the reach of near-future experiments are collected for example in Ref.~\cite{Schwingenheuer:2012zs}. The current best upper bounds on $|m_{ee}|$ are in the range $|m_{ee}|<0.12\text{--}0.2~\text{eV}$, depending on uncertainties in the nuclear matrix elements. The KamLAND-Zen (KLZ) experiment obtained a 90\%-CL lower bound $T^{0\nu}_{1/2}(^{136}{\rm Xe})>1.9\times10^{25}$ yr on the $0\nu\beta\beta$-decay half-life of $^{136}{\rm Xe}$~\cite{Gando:2012zm}. The EXO-200 (EXO) experiment reported a 90\%-CL lower limit $T^{0\nu}_{1/2}(^{136}{\rm Xe})>1.6\times10^{25}$ yr ~\cite{Auger:2012ar}. Combining the KLZ and EXO bounds gives $T^{0\nu}_{1/2}(^{136}{\rm Xe})>3.4\times10^{25}$ yr at the 90\% CL, which corresponds to an upper limit $|m_{ee}|<0.120-0.250$ eV (once account is taken of the uncertainties in the available nuclear matrix elements). The GERDA experiment~\cite{Agostini:2013mzu} in its phase I has published a new limit on the $^{76}{\rm Ge}$ $0\nu\beta\beta$-decay half-life $T^{0\nu}_{1/2}(^{76}{\rm Ge})>2.1\times10^{25}$ yr at the 90\% CL. Combining it with the previous Ge-based results (Heidelberg-Moscow~\cite{KlapdorKleingrothaus:2000sn} and IGEX~\cite{Aalseth:2004wf}) yields $T^{0\nu}_{1/2}(^{76}{\rm Ge})>3.0\times10^{25}$ yr at $90\%$ CL. This corresponds to $|m_{ee}|<0.20-0.40$ eV.

We mention here in passing that the phase-I GERDA limit excludes the $^{76}{\rm Ge}$ $0\nu\beta\beta$-decay signal claimed in Ref.~\cite{KlapdorKleingrothaus:2006ff} with a half-life $T^{0\nu}_{1/2}(^{76}{\rm Ge})=2.23^{+0.44}_{-0.31}\times10^{25}$ yr at the 68\% CL, independently of uncertainties in the nuclear matrix elements and of the physical mechanism responsible for $0\nu\beta\beta$-decay. The KLZ and EXO results exclude the claim in \cite{KlapdorKleingrothaus:2006ff} at more than 97.5\% CL, but the comparison is model dependent.

In the near future, KamLAND-Zen, EXO, and GERDA are expected to probe the range
\begin{eqnarray}
  0.01~\text{eV}<|m_{ee}|<0.1~\text{eV}.
  \label{meeSensitivity}
\end{eqnarray}
If these experiments measure a value of $|m_{ee}|>0.01$ eV, the hierarchical spectrum of normal mass ordering would be strongly disfavored~\cite{Bilenky:2001rz}.

\begin{figure}[t]
\begin{minipage}[t]{7.5cm}
\epsfig{figure=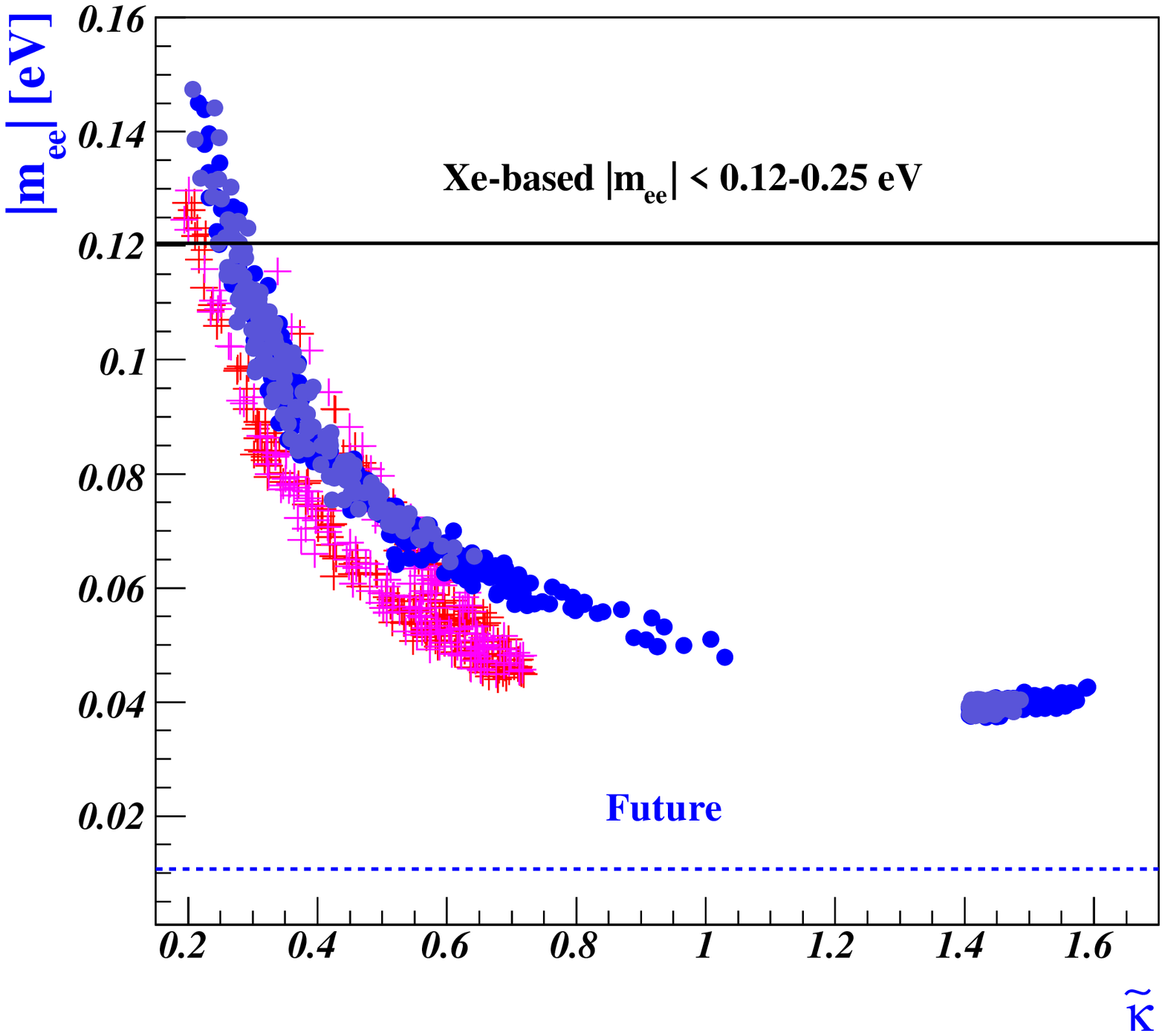,width=7.5cm,angle=0}
\end{minipage}
\hspace*{1.0cm}
\begin{minipage}[t]{7.5cm}
\epsfig{figure=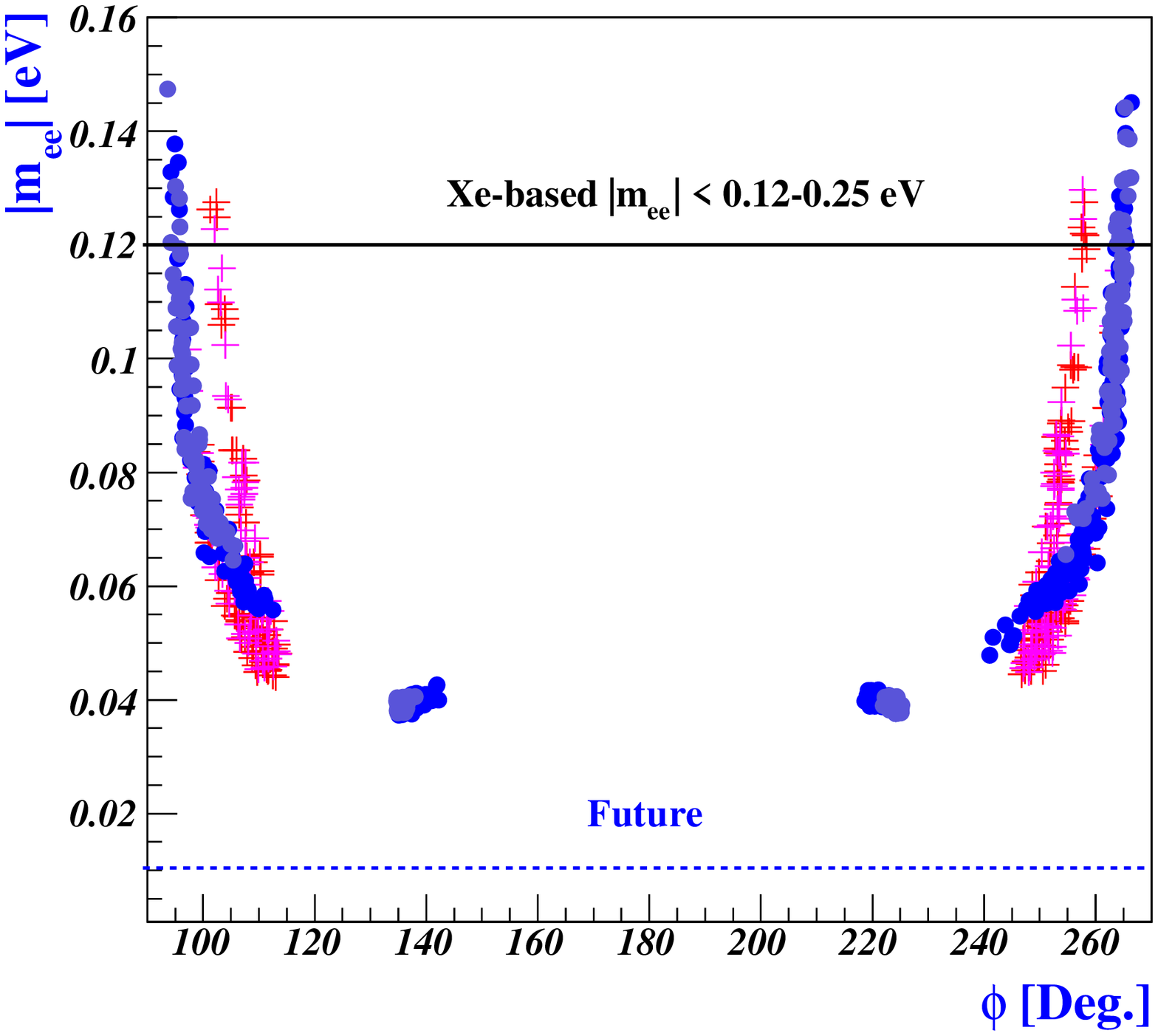,width=7.5cm,angle=0}
\end{minipage}
\caption{\label{FigA3} Plots of $|m_{ee}|$ as a function of $\tilde{\kappa}$ (left) and $\phi$ (right). The horizontal solid (dotted) lines show the current bounds from (near future reach of) Xe-based $0\nu\beta\beta$ experiments.}
\end{figure}

\begin{figure}[t]
\begin{minipage}[h]{7.5cm}
\epsfig{figure=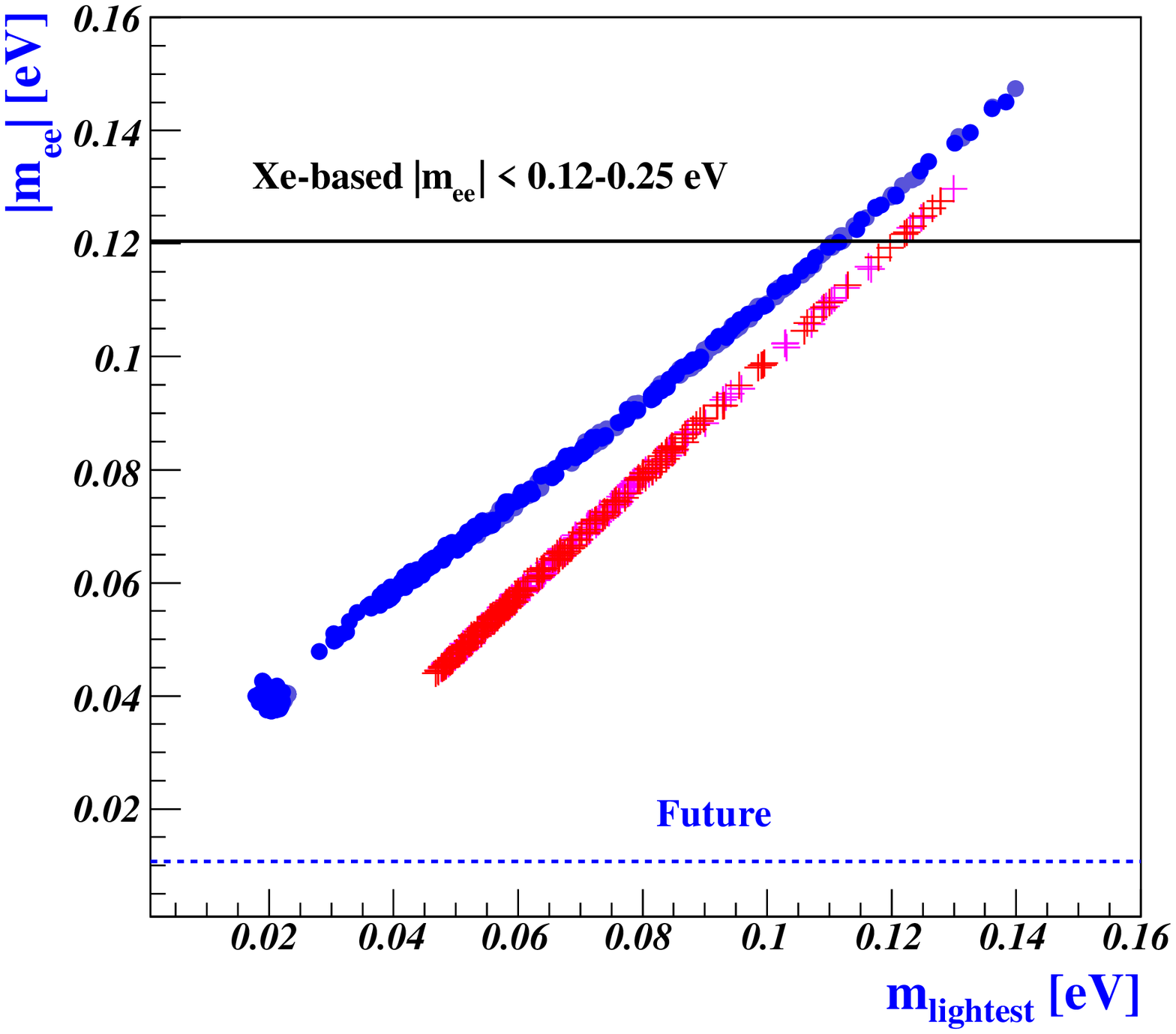,width=7.5cm,angle=0}
\end{minipage}
\hspace*{1.0cm}
\begin{minipage}[h]{7.5cm}
\epsfig{figure=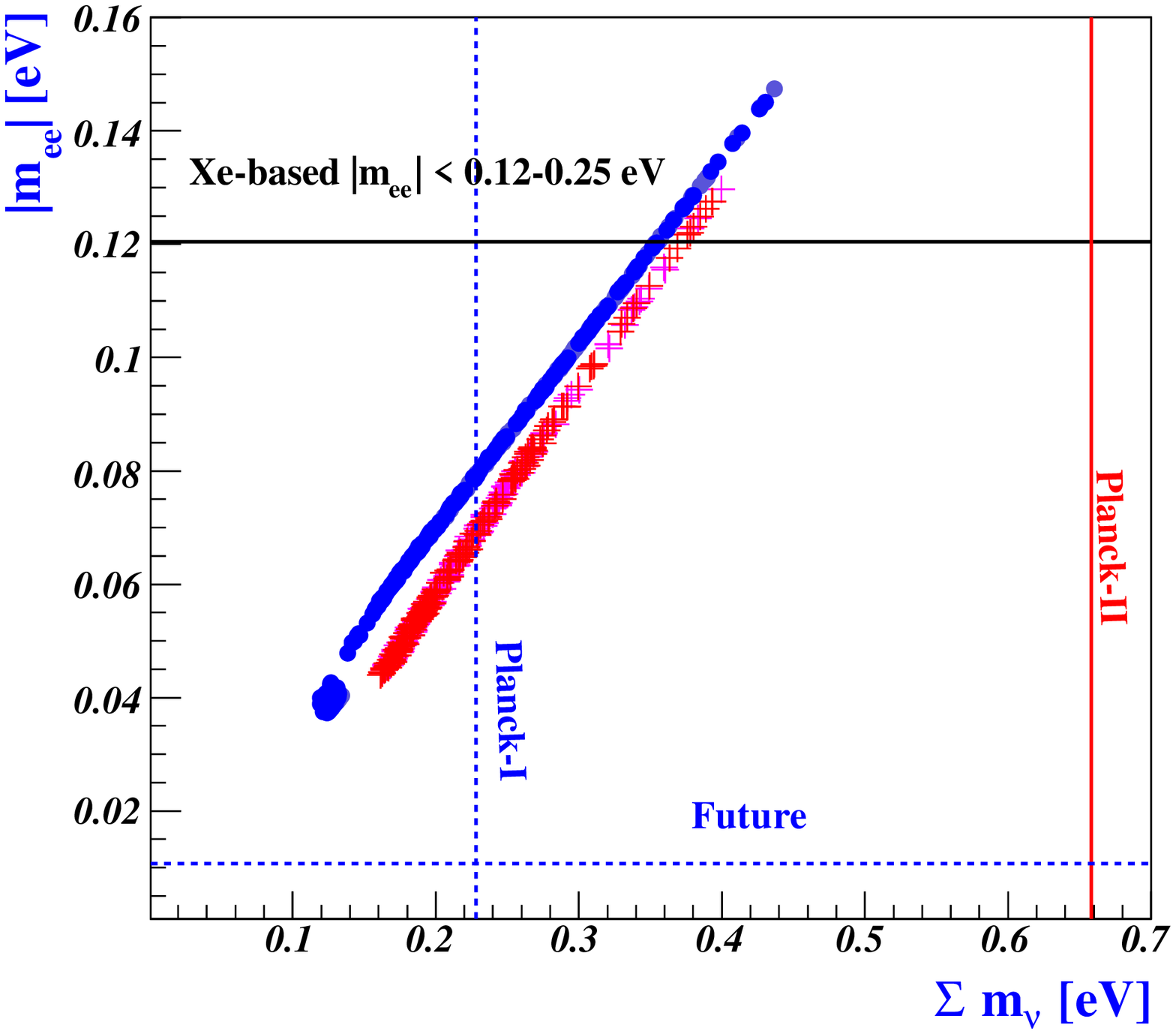,width=7.5cm,angle=0}
\end{minipage}
\caption{\label{FigA2} Plots of $|m_{ee}|$ as a function of $m_{\rm lightest}$ (left) and $\sum m_{i}$ (right). The horizontal solid (dotted) lines show the current bounds from (near future reach of) Xe-based $0\nu\beta\beta$ experiments, while the vertical solid (dotted) lines indicate the cosmological Planck-I (Planck-II) upper bounds.}
\end{figure}

In our model, the effective neutrino mass $|m_{ee}|$ that characterizes the amplitude for $0\nu\beta\beta$-decay is given by
 \begin{eqnarray}
  |m_{ee}|= m_{0}\left|\frac{3+\tilde{\kappa}e^{i\phi}}{1+\tilde{\kappa}e^{i\phi}}\right|~.
  \label{mee1}
 \end{eqnarray}
This shows that in our model the rate of $0\nu\beta\beta$-decay depends on the parameters $m_{0}$, $\tilde{\kappa}$, and $\phi$ associated with the heavy (right-handed) Majorana neutrinos in Eq.~(\ref{Yuk:MR}). These are the same parameters that enter leptogenesis~\cite{review}.

Varying our model parameters within the $3\sigma$ experimental bounds of Table~\ref{exp} produces the results shown in Figs.~\ref{FigA3} and~\ref{FigA2}. The horizontal solid (dotted) lines provide a rough indication of the current Xe-based upper bounds (near-future reach) of $0\nu\beta\beta$ experiments. Fig.~\ref{FigA3} shows the sensitivity of $|m_{ee}|$ to the input parameters $\tilde{\kappa}$ (left plot) and $\phi$ (right plot).
In Fig.~\ref{FigA2}, the plot on the left shows the dependence of $|m_{ee}|$ on the lightest neutrino mass $m_{\rm lightest}$, which equals $m_{1}$ for NO and $m_{3}$ for IO. The plot on the right shows $|m_{ee}|$ vs.\ the sum of the light neutrino masses $\sum_{i=1}^{3} m_{i}$, which is subject to the cosmological bounds indicated by the vertical solid and dotted lines. These bounds are $\sum_{i}m_{i}<0.23$ eV at $95\%$ CL (Planck-I, derived from the combination Planck + WMAP low-multipole polarization + high resolution CMB + baryon acoustic oscillations (BAO), assuming a standard $\Lambda$CDM cosmological model) and $\sum_{i}m_{i}<0.66$ eV at $95\%$ CL (Planck-II, derived from the data without BAO~\cite{Ade:2013zuv}). The more stringent Planck I limit cuts into our region of points and starts to disfavor a quasi-degenerate light neutrino mass spectrum.
The current $0\nu\beta\beta$-decay experiments also cut into our region of points, and the near-future $0\nu\beta\beta$-decay experiments can test our model completely.

We conclude this section by remarking that the tritium beta decay experiment KATRIN~\cite{KATRIN} is not expected to reach into our model region. KATRIN will be sensitive to an effective electron neutrino mass  $m_{\beta}=\sqrt{\sum_{i}|U_{ei}|^{2}\,m^2_{i}}$~\cite{beta} down to about $0.2$ eV, while our model produces values in the range $0.047\lesssim m_{\nu_e}\lesssim0.130$ eV for NO and $0.049\lesssim m_{\nu_e}\lesssim0.150$ eV for IO.

\begin{figure}[t]
\begin{minipage}[h]{7.5cm}
\epsfig{figure=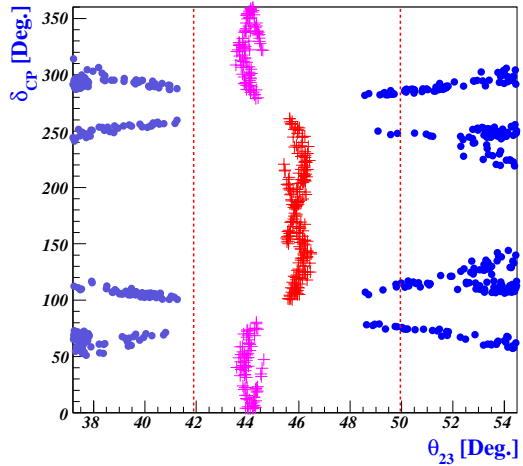,width=7.5cm,angle=0}
\end{minipage}
\caption{\label{FigA4} Predictions for the leptonic CP-violating $\delta_{CP}$ as a function of neutrino mixing angle $\theta_{23}$. The vertical dotted lines bound the current best-fit value of $\theta_{23}$. The blue dots and the red crosses correspond to the inverted mass ordering (IO) and the normal mass ordering (NO), respectively.}
\end{figure}
\subsection{Leptonic CP violation}
After the observation of a non-zero mixing angle $\theta_{13}$ in the Daya Bay~\cite{An:2012eh} and RENO~\cite{Ahn:2012nd} experiments, the Dirac CP-violating phase $\delta_{CP}$ is the next observable on the agenda of neutrino oscillation experiments. The magnitude of the CP-violating effects is determined by the invariant $J_{CP}$ associated with the Dirac CP-violating phase:
 \begin{eqnarray}
 J_{CP}\equiv-{\rm Im}[U^{\ast}_{e1}U_{e3}U_{\tau1}U^{\ast}_{\tau3}]=\frac{1}{8}\sin2\theta_{12}\sin2\theta_{13}\sin2\theta_{23}\cos\theta_{13}\sin\delta_{CP}~.
 \label{JCP}
 \end{eqnarray}
Here $U_{\alpha j}$ is an element of the PMNS matrix in Eq.~(\ref{rebasing1}), with $\alpha=e,\mu,\tau$
corresponding to the lepton flavors and $j=1,2,3$ corresponding to the light neutrino mass eigenstates.

Due to the precise measurement of $\theta_{13}$, which is relatively large, it may now be possible to put constraints on the Dirac phase $\delta_{CP}$ which will be obtained in the long baseline  neutrino oscillation experiments T2K, NO$\nu$A, etc. (see, Ref.~\cite{PDG}). However, the current large uncertainty on $\theta_{23}$ is at present limiting the information that can be extracted from the $\nu_{e}$ appearance measurements. Precise measurements of all the mixing angles are needed to maximize the sensitivity to the leptonic CP violation.

Fig.~\ref{FigA4} shows our model predictions for the Dirac CP-violating phase $\delta_{CP}$ in terms of the atmospheric mixing angle $\theta_{23}$. The blue dots and red crosses correspond to the inverted mass ordering (IO) and the normal mass ordering (NO), respectively. Within our model, future precise measurements of $\theta_{23}$ should be able to distinguish between IO and NO. For NO, $\theta_{23}$ would be in the range $[43^\circ,47^\circ]$, close to the maximal value of $45^\circ$. For IO, $\theta_{23}$ would be in the range $[37^\circ,41^\circ] \cup [49^\circ,54^\circ]$, that is $5^\circ$ to $8^\circ$ away from maximality. In turn, such precise measurements of $\theta_{23}$ would restrict the possible range of $\delta_{CP}$ in our model. A value of $\theta_{23}$ slightly larger than maximal, i.e.\ $\theta_{23} \in [45^\circ,47^\circ]$, would imply an NO and $\delta_{CP} \in [90^\circ,270^\circ]$, while a value of $\theta_{23}$ slightly smaller than maximal, i.e.\ $\theta_{23} \in [43^\circ,45^\circ]$, would imply an NO and $\delta_{CP} \in [0,90^\circ] \cup [270^\circ,360^\circ]$. A value of $\theta_{43}$ considerably larger or smaller than maximal, i.e.\ $[37^\circ,41^\circ] \cup [49^\circ,54^\circ]$, would imply IO and $\delta_{CP}$ within few degrees of $70^\circ$, $110^\circ$, $250^\circ$, or $290^\circ$.

\begin{figure}[b]
\begin{minipage}[t]{7.5cm}
\epsfig{figure=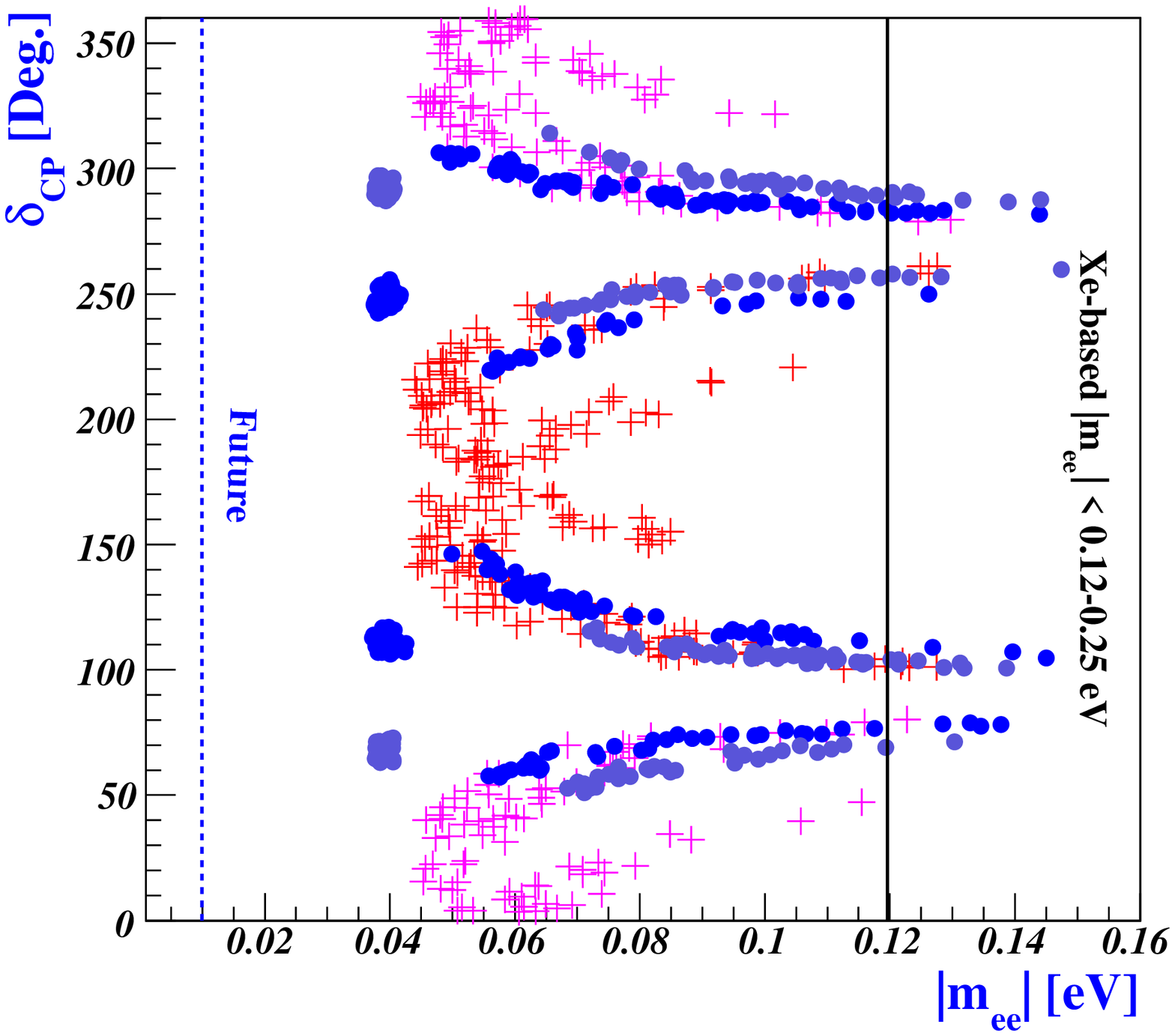,width=7.5cm,angle=0}
\end{minipage}
\hspace*{1.0cm}
\begin{minipage}[t]{7.5cm}
\epsfig{figure=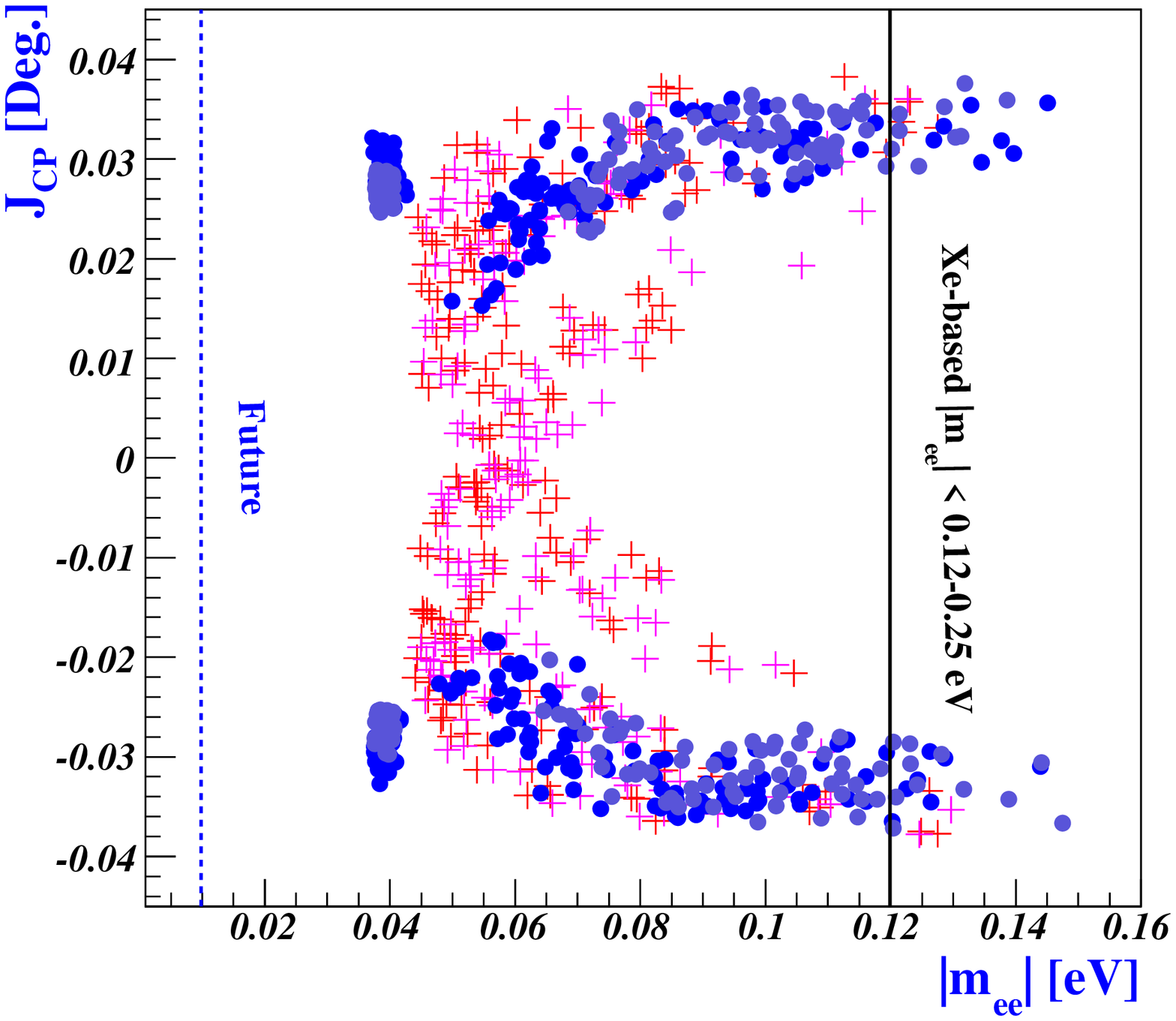,width=7.5cm,angle=0}
\end{minipage}
\caption{\label{FigA5} Plots of the leptonic CP-violating phase $\delta_{CP}$ (left) and the CP-violating invariant $J_{CP}$ (right) vs.\ $|m_{ee}|$. The vertical solid (dotted) lines indicate the current bounds from (near-future reach of) Xe-based $0\nu\beta\beta$-decay experiments.}
\end{figure}

In our model, the magnitudes of the CP-violating quantities $J_{CP}$ and $\delta_{CP}$ are constrained by the neutrino mass matrix Eq.~(\ref{meff}) and depend on the value of the phase $\phi$.
The Jarlskog invariant $J_{CP}$ can be expressed in terms of the elements of the matrix $h=\mathcal{M}_{\nu}\mathcal{M}^{\dag}_{\nu}$ as~\cite{Branco:2002xf}
 \begin{eqnarray}
  J_{CP}=-\frac{{\rm Im}\{h_{12}h_{23}h_{31}\}}{\Delta m^{2}_{21}\Delta m^{2}_{31}\Delta m^{2}_{32}}~.
  \label{JCP2}
 \end{eqnarray}
In our model, the numerator is expressed as
 \begin{eqnarray}
  {\rm Im}\{h_{12}h_{23}h_{31}\}&=&\sin\phi \, m^{6}_{0}\frac{27y^{2}_{2}y^{2}_{3}\tilde{\kappa}^3(y^{2}_{3}-y^{2}_{2})}{[(1+\tilde{\kappa}^{2})^2-4\tilde{\kappa}^2\cos^2\phi]^2}\Big\{2(1-y^{2}_{2})(1-y^{2}_{3})+\tilde{\kappa}^2(2+y^{2}_{2}y^{2}_{3})\nonumber\\
  &+&2\{y^{2}_{2}(2+y^{2}_{3})-2(1-y^{2}_{3})\}\tilde{\kappa}\cos\phi\Big\}~.
  \label{JCP1}
 \end{eqnarray}
Clearly, Eq.~(\ref{JCP1}) indicates that $J_{CP}$ depends on the phase $\phi$, and in the limits $y_{2}\rightarrow y_{3}$ or $\sin\phi\rightarrow0$, the leptonic CP-violating invariant $J_{CP}$ goes to zero.


The dependence of $\delta_{CP}$ and $J_{CP}$ on the effective Majorana neutrino mass $|m_{ee}|$ is shown in Fig.~\ref{FigA5}.
The left plot shows predictions for $\delta_{CP}$, the right plot for $J_{CP}$. The vertical solid (dotted) lines show the current bounds from (near future reach of) Xe-based $0\nu\beta\beta$-decay experiments. The correlations shown in the figure indicate that  in our model precise measurements of or improved upper bounds on $|m_{ee}|$ from $0\nu\beta\beta$-decay experiments may be able to restrict the possible ranges of $\delta_{CP}$, and in some cases may even distinguish NO from IO.

It is worth remarking that in the context of our model an observation of $0\nu\beta\beta$-decay and an accurate measurement of its half-life, combined with data on the absolute neutrino mass scale, may be able to provide information on the Majorana phases in the PMNS matrix.
Similarly to Eq.~(\ref{JCP}), two CP-violating invariants can be defined in place of the Majorana phases $\varphi_{1,2}$~\cite{Jenkins:2007ip},
 \begin{eqnarray}
 J^{Mj_1}_{CP}&\equiv&{\rm Im}[U^{2}_{e2}(U^{\ast}_{e1})^2]=\frac{1}{4}\sin^2\theta_{12}\cos^{4}\theta_{13}\sin(\varphi_1-\varphi_2)~,\nonumber\\
 J^{Mj_2}_{CP}&\equiv&{\rm Im}[U^{2}_{e3}(U^{\ast}_{e1})^2]=\frac{1}{4}\sin^2\theta_{13}\cos^{2}\theta_{12}\sin(\varphi_1-2\delta_{CP})~.
 \label{JCPMj}
 \end{eqnarray}
In the parametrization of the PMNS matrix in Eq.~(\ref{rebasing1}), the Majorana CP phases can be extracted as
 $\varphi_{1}=2\arg\left(U^{\ast}_{e1}\right)$, $\varphi_{2}=2\arg\left(U^{\ast}_{e2}\right)$.
Since there is no distinction between the $0\nu\beta\beta$ rate of a nucleus and that of its antinucleus, $0\nu\beta\beta$-decay processes do not exhibit CP violation~\cite{Barger:2002vy}. There are, however, processes that do manifest CP-violating effects and that can be sensitive to the CP violation induced by the Majorana phases $\varphi_1$ and $\varphi_2$~\cite{deGouvea:2002gf}: (i) neutrino $\leftrightarrow$ antineutrino oscillations~\cite{Xing:2013woa}, (ii) rare leptonic decays of $K$ and $B$ mesons, such as $K^{\pm}\rightarrow\pi^{\mp}l^{\pm}l^{\pm}$ and similar modes for the $B$, and (iii) leptogenesis in the early Universe.

\subsection{Leptogenesis}
Baryogenesis through leptogenesis is governed by the same CP-violating phases that enter the quantities $|m_{ee}|$ and $J_{CP}$. It is therefore interesting to ask if the parameters that produce a correct baryon asymmetry parameter $\eta_{B}$ also provide sizable values of $|m_{ee}|$ and/or $\delta_{CP}$.\footnote{Since there exists a unique CP phase in the model, Majorana CP phases can also be linked to directly $\eta_{B}$.}

\begin{figure}[h]
\begin{minipage}[t]{7.5cm}
\epsfig{figure=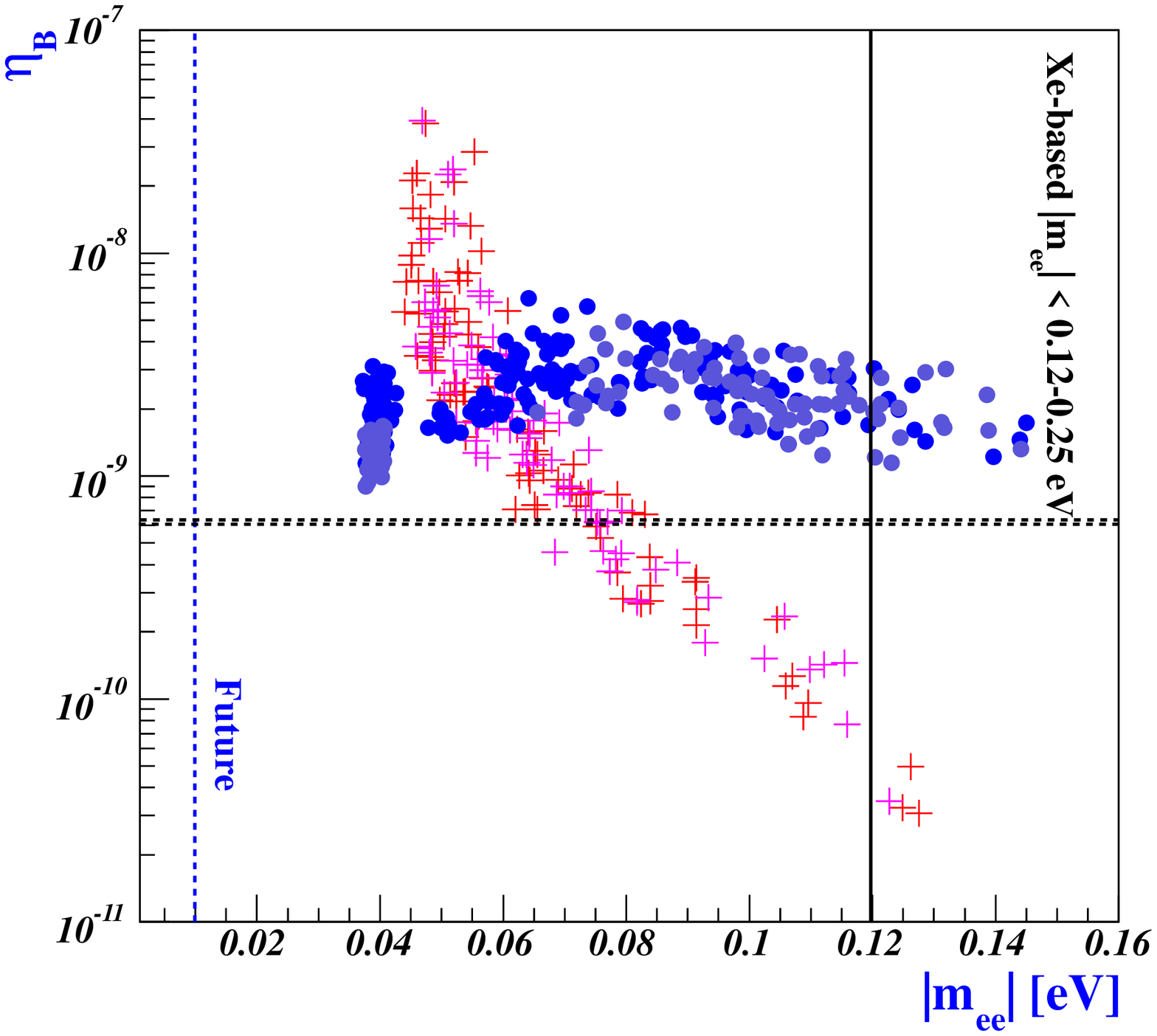,width=7.5cm,angle=0}
\end{minipage}
\hspace*{1.0cm}
\begin{minipage}[t]{7.5cm}
\epsfig{figure=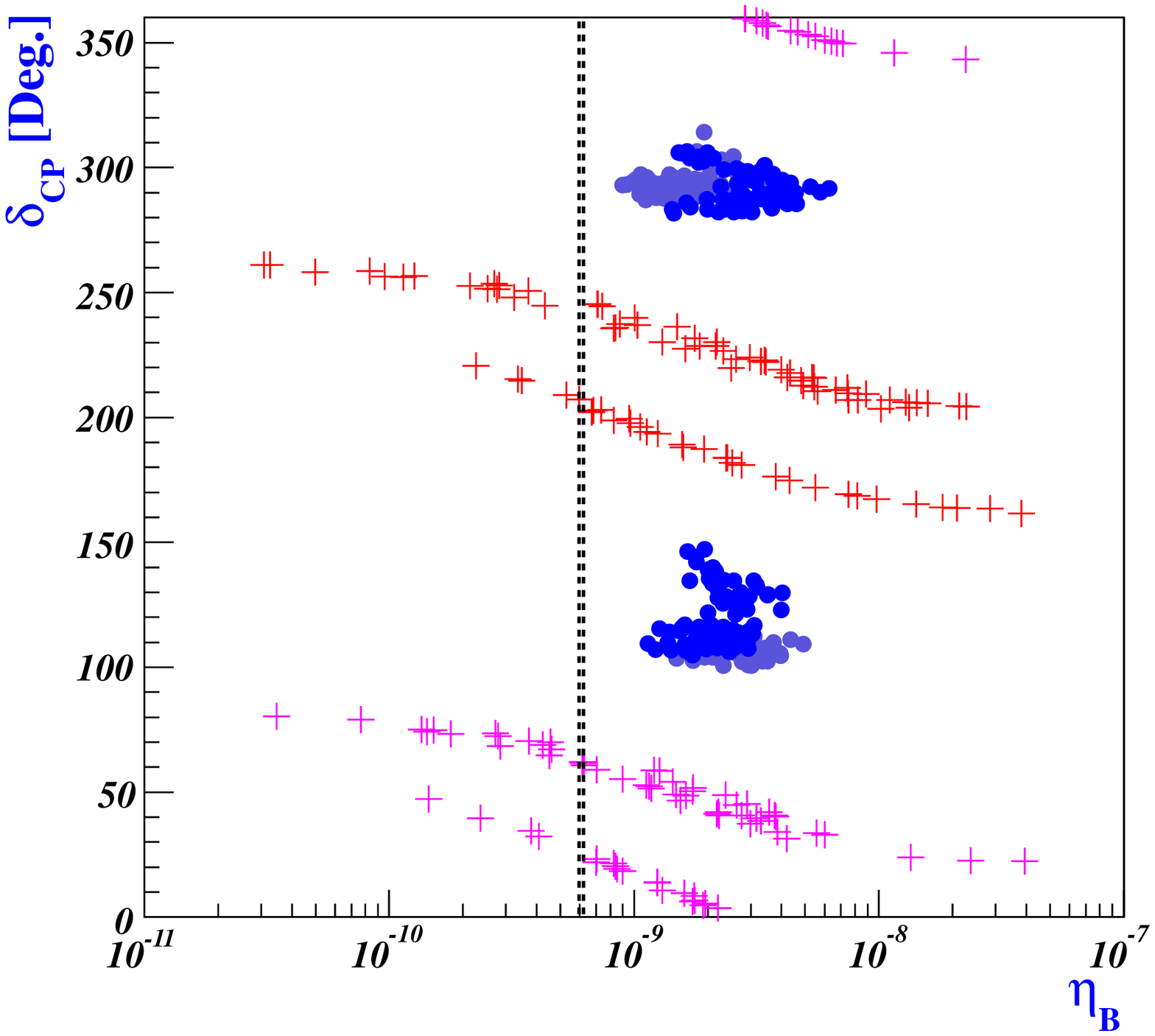,width=7.5cm,angle=0}
\end{minipage}
\caption{\label{FigA6} Model predictions for primordial baryogenesis. The plot on the left shows the relation between the baryon asymmetry parameter $\eta_{B}$ and the $0\nu\beta\beta$-decay mass $|m_{ee}|$. The vertical solid line indicates the current bounds from Xe-based $0\nu\beta\beta$-decay experiments, while the dotted line indicates their near-future reach. The plot on the right shows our model predictions for $\delta_{CP}$ in terms of the positive values of $\eta_{B}$. The thick dashed line on both plots corresponds to the measured values of the baryon asymmetry in the Universe $\eta_{B}=(6.05\pm 0.07)\times 10^{-10}$ from Planck measurements~\cite{Ade:2013zuv}, or $\eta_{B}=(6.2\pm 1.0)\times 10^{-10}$ from D/H measurements~\cite{Cooke:2013}.}
\end{figure}
Leptogenesis in the early universe is expected to occur at an energy scale where the $A_{4}$ symmetry is broken but the SM gauge group remains unbroken. Since the Dirac neutrino Yukawa couplings are $\approx0.1$, the scale of leptogenesis corresponds to $\approx10^{12}$ GeV, and flavorless leptogenesis is viable.
The CP asymmetry is generated through the interference between tree and one-loop diagrams for the decay of the $i$-th generation
heavy Majorana neutrino $N_{i}$ into $\Phi$ and leptons~\cite{review}. This decay rate is given by the expression
 \begin{eqnarray}
  \varepsilon_{i}=\frac{|y^{\nu}_{1}|^{2}v^{2}_{\Phi}\left\{-P_{i}g(x_{ij})+P_{k}g(x_{ik})\right\}}{16\pi (\tilde{m}^{\dag}_{D}\tilde{m}_{D})_{ii}},~
 \end{eqnarray}
where $(i,j,k)=(1,2,3)$ and cyclic permutations, $P_{1}=(2-y^{2}_{2}-y^{2}_{3})^2(1+\tilde{\kappa}\cos\phi)/a_{+}$, $P_{2}=3(y^{2}_{2}-y^{2}_{3})^2(1-\tilde{\kappa}\sin\phi)/a_{-}$, $P_{3}=3(y^{2}_{2}-y^{2}_{3})^2\tilde{\kappa}\sin\phi/{a_{+}a_{-}}$, where $a_{\pm}=\sqrt{1+\tilde{\kappa}^{2}\pm 2\tilde{\kappa}\cos\phi}$, and $g(x)$ is a loop function defined by
\begin{eqnarray}  g(x_{ij})=\sqrt{x_{ij}}\left(\frac{1}{1-x_{ij}}+1-(1+x_{ij})\ln\left(\frac{1+x_{ij}}{x_{ij}}\right)\right)\,, \end{eqnarray}
with $x_{ij}=M^{2}_{j}/M^{2}_{i}$.
Moreover, $\tilde{m}_{D}=m_{D}U_{R}$, where
\begin{eqnarray}
  U_{R} = \frac{1}{\sqrt{2}}{\left(\begin{array}{ccc}
  0  &  \sqrt{2}  &  0 \\
  1 &  0  &  -1 \\
  1 &  0  &  1
  \end{array}\right)}{\left(\begin{array}{ccc}
  e^{i\frac{\psi_1}{2}}  &  0  &  0 \\
  0  &  1  &  0 \\
  0  &  0  &  e^{i\frac{\psi_2}{2}}
  \end{array}\right)},
\end{eqnarray}
with $\psi_1 = \tan^{-1} \!\Big( \frac{-\tilde{\kappa}\sin\phi}{1+\tilde{\kappa}\cos\phi} \Big)$ and $\psi_2 = \tan^{-1} \!\Big( \frac{\tilde{\kappa}\sin\phi}{1-\tilde{\kappa}\cos\phi} \Big)$.

In the limit $y_{2,3}\rightarrow1$, the CP-violating quantities $J_{CP}$ and $\varepsilon_{i}$ vanish.
Near this limit, the cosmological baryon asymmetry is given by~\cite{review}:
 \begin{eqnarray}
  \eta_{B}\simeq
  -0.01\sum_{i}\varepsilon_{i}\,\tilde{\kappa}(\tilde{m}_{i})~,
 \end{eqnarray}
where $\tilde{\kappa}(\tilde{m}_{i})$ is a wash-out factor given approximately by $\tilde{\kappa}(\tilde{m}_{i})\simeq\big[(8.25/\tilde{m}_{i})+(\tilde{m}_{i}/0.2)^{1.16}\big]^{-1}$,
with $\tilde{m}_{i}=(\tilde{m}^{\dag}_{D}\tilde{m}_{D})_{ii}/M_{i}$ in meV~\cite{lepto2}.

Fig.~\ref{FigA6} shows the values of the baryon asymmetry parameter $\eta_B$ in our model vs.\ the $0\nu\beta\beta$-decay mass $|m_{ee}|$ and the CP-violating phase $\delta_{CP}$. The plot on the left shows positive values of $\eta_{B}$ in terms of $|m_{ee}|$. The plot on the right shows predictions of $\delta_{CP}$ as functions of positive values of $\eta_{B}$. Observationally, $\eta_{B}=(6.05\pm 0.07)\times 10^{-10}$ from Planck measurements~\cite{Ade:2013zuv}, or $\eta_{B}=(6.2\pm 1.0)\times 10^{-10}$ from D/H measurements~\cite{Cooke:2013}.  In Fig.~\ref{FigA6}, these values (almost indistinguishable at the scale of the plots) are indicated by a thick dashed line.

Our model is compatible with a successful baryogenesis through leptogenesis scenario. Imposing a successful leptogenesis constrains both the Dirac CP-violating phase (or $J_{CP}$) and the rate of $0\nu\beta\beta$-decays. In correspondence to the observational values of $\eta_B$, a successful leptogenesis in our model requires a normal mass ordering (NO), fixes a Dirac CP-violating phase equal to approximately one of the four values $20^\circ$, $60^\circ$, $205^\circ$, or $245^\circ$ (the first two values correspond to $\theta_{23}\simeq44^{\circ}$ and the last two values to $\theta_{23}\simeq46^{\circ}$), and constrains the $0\nu\beta\beta$ Majorana mass to be $|m_{ee}| \simeq 0.072\pm 0.012$~eV. Also, the mass of the lightest neutrino would be $\simeq 0.07$ eV, and the sum of the light neutrino masses would be $\sum m_i \simeq 0.22$ eV, which is reachable with upcoming cosmological measurements~\cite{Amendola:2012ys}.
Note that since the magnitude of the Dirac neutrino Yukawa couplings is $\approx {\cal O}(0.1)$, due to the seesaw relation $y^{\nu2}_{1}=2Mm_{0}/v^2_{\Phi}$ in Eq.~(\ref{meff}), the leptogenesis scale in our model lies approximately in the range $\approx10^{12}-10^{14}$ GeV.

\section{Conclusions}
We have proposed an economical model based on $SU(2)_{L}\times U(1)_{Y}\times A_{4}\times U(1)_X$ in a seesaw framework, in which the Yukawa couplings are functions of flavon fields that decouple at some large flavor physics scale. By appropriate assignments of $U(1)_X$ charges to the  quark and lepton flavors, our model can naturally explain the mass hierarchies and the pattern of mixing angles in both the quark and lepton sectors: two large and one small mixing angles for the quarks; light neutrinos, one large and two small mixing angles for the leptons. An important point is that our model shows why the hierarchy of light neutrino masses is mild, while the hierarchy of the charged fermions is strong.

\begin{figure}[t]
\begin{minipage}[h]{7.5cm}
\epsfig{figure=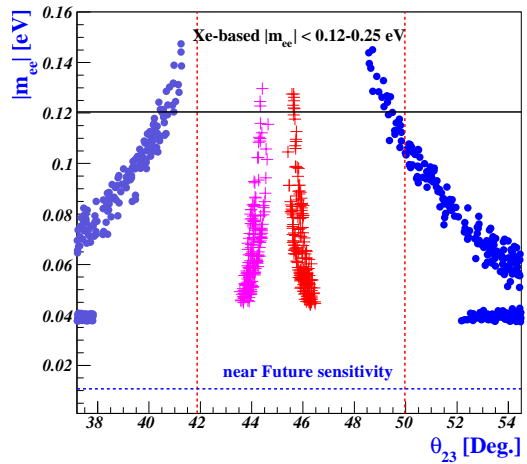,width=7.5cm,angle=0}
\end{minipage}
\caption{\label{FigA7} Model prediction of $|m_{ee}|$ in terms of $\theta_{23}$. The vertical dotted lines show the best-fit values for $\theta_{23}$. The horizontal lines show the current upper bounds from (and near-future reach of) $0\nu\beta\beta$-decay experiments. Blue dots correspond to the inverted mass ordering (NO) and red crosses to the normal mass ordering (IO).}
\end{figure}

Our model predictions for the yet unmeasured leptonic CP-violating phase $\delta_{CP}$ and the neutrinoless $\beta\beta$-decay effective mass $|m_{ee}|$ can be fully tested in current and upcoming experiments.
For both normal and inverted mass orderings in the neutrino masses,
the allowed regions of $|m_{ee}|$ and $\theta_{23}$ in our model are strongly restricted and they are accessible in $0\nu\beta\beta$-decay experiments (such as GERDA-II, MAJORANA, CUORE, and others listed in Ref.~\cite{Schwingenheuer:2012zs}) and long-baseline neutrino oscillation experiments (such as T2K, NO$\nu$A, and others listed in Ref.~\cite{PDG}).

Future precise measurements of $|m_{ee}|$ and $\theta_{23}$ are also able in principle to exclude or favor our model, as summarized in Fig.~\ref{FigA7}. There we plot our model predictions for the correlation between $|m_{ee}|$ and $\theta_{23}$. For the normal mass ordering, our model predicts that $\theta_{23}$ must be within $\sim1^{\circ}$ of $45^\circ$. For the inverted mass ordering, our model predicts that $\theta_{23}$ must be $\sim4^\circ$ to $\sim8^\circ$ degrees away from $45^\circ$ (in either direction). For both normal and inverted mass ordering, our model predicts that $0.035\text{eV}<|m_{ee}|\lesssim0.15$ eV.

Finally, with flavon Dirac neutrino Yukawa couplings $\hat{y}\approx 0.1$, our model predicts values of $|m_{ee}|$, $\delta_{CP}$, and the atmospheric mixing angle $\theta_{23}$ that can accommodate a successful leptogenesis in the early universe. This happens for a $0\nu\beta\beta$-decay mass $|m_{ee}|\simeq0.072\pm0.012$ eV, and a Dirac CP-violating phase $\delta_{CP}$ equal to either $\delta_{CP} \simeq20^{\circ}$ or $60^{\circ}$ (for $\theta_{23}\simeq44^{\circ}$) or $\delta_{CP} \simeq 205^{\circ}$ or $245^{\circ}$ (for $\theta_{23}\simeq46^{\circ}$).

\newpage
\appendix
\section{The $A_{4}$ Group}
 \label{A4group}
The  group $A_{4}$ is the symmetry group of the
tetrahedron, isomorphic to the finite group of the even permutations of four
objects. The group $A_{4}$ has two generators, denoted $S$ and $T$, satisfying the relations $S^{2}=T^{3}=(ST)^{3}=1$. In the three-dimensional real representation, $S$ and $T$ are given by
 \begin{eqnarray}
 S={\left(\begin{array}{ccc}
 1 &  0 &  0 \\
 0 &  -1 & 0 \\
 0 &  0 &  -1
 \end{array}\right)}~,\qquad T={\left(\begin{array}{ccc}
 0 &  1 &  0 \\
 0 &  0 &  1 \\
 1 &  0 &  0
 \end{array}\right)}~.
 \label{generator}
 \end{eqnarray}
$A_{4}$ has four irreducible representations: one triplet ${\bf
3}$ and three singlets ${\bf 1}, {\bf 1}', {\bf 1}''$. An $A_4$ triplet $(a_1,a_2,a_3)$ transforms in the unitary representation by multiplication with the $S$ and $T$ matrices in Eq.~(\ref{generator}) above,
\begin{align}
S \begin{pmatrix} a_1 \\ a_2 \\ a_3 \end{pmatrix} = \begin{pmatrix} a_1 \\ -a_2 \\ -a_3 \end{pmatrix},
\qquad
T \begin{pmatrix} a_1 \\ a_2 \\ a_3 \end{pmatrix} = \begin{pmatrix} a_2 \\ a_3 \\ a_1 \end{pmatrix}.
\end{align}
An $A_4$ singlet $a$ is invariant under the action of $S$ ($Sa=a$), while the action of $T$ produces $Ta=a$ for ${\bf 1}$, $Ta=\omega a$ for ${\bf 1}'$, and $Ta=\omega^2 a$ for ${\bf 1}''$, where $\omega=e^{i2\pi/3}=-1/2+i\sqrt{3}/2$ is a complex cubic-root of unity.
Products of two $A_4$ representations decompose into irreducible representations according to the following multiplication rules: ${\bf 3}\otimes{\bf 3}={\bf 3}_{s}\oplus{\bf
3}_{a}\oplus{\bf 1}\oplus{\bf 1}'\oplus{\bf 1}''$, ${\bf
1}'\otimes{\bf 1}''={\bf 1}$, ${\bf 1}'\otimes{\bf 1}'={\bf 1}''$
and ${\bf 1}''\otimes{\bf 1}''={\bf 1}'$. Explicitly, if $(a_{1},
a_{2}, a_{3})$ and $(b_{1}, b_{2}, b_{3})$ denote two $A_4$ triplets,
 \begin{eqnarray}
  (a\otimes b)_{{\bf 3}_{\rm s}} &=& (a_{2}b_{3}+a_{3}b_{2}, a_{3}b_{1}+a_{1}b_{3}, a_{1}b_{2}+a_{2}b_{1})~,\nonumber\\
  (a\otimes b)_{{\bf 3}_{\rm a}} &=& (a_{2}b_{3}-a_{3}b_{2}, a_{3}b_{1}-a_{1}b_{3}, a_{1}b_{2}-a_{2}b_{1})~,\nonumber\\
  (a\otimes b)_{{\bf 1}} &=& a_{1}b_{1}+a_{2}b_{2}+a_{3}b_{3}~,\nonumber\\
  (a\otimes b)_{{\bf 1}'} &=& a_{1}b_{1}+\omega^{2} a_{2}b_{2}+\omega a_{3}b_{3}~,\nonumber\\
  (a\otimes b)_{{\bf 1}''} &=& a_{1}b_{1}+\omega a_{2}b_{2}+\omega^{2} a_{3}b_{3}~.
 \end{eqnarray}

To make the presentation of our model physically more transparent, we define the $T$-flavor quantum number $T_f$ through the eigenvalues of the operator $T$, for which $T^3=1$. In detail, we say that a field $f$ has $T$-flavor $T_f=0$, +1, or -1 when it is an eigenfield of the $T$ operator with eigenvalue $1$, $\omega$, $\omega^2$, respectively (in short, with eigenvalue $\omega^{T_f}$ for $T$-flavor $T_f$, considering the cyclical properties of the cubic root of unity $\omega$). The $T$-flavor is an additive quantum number modulo 3. We also define the $S$-flavor-parity through the eigenvalues of the operator $S$, which are +1 and -1 since $S^2=1$, and we speak of $S$-flavor-even and $S$-flavor-odd fields.
For $A_4$-singlets, which are all $S$-flavor-even, the representation $\mathbf{1}$ is $T$-flavorless ($T_f=0$), the representation $\mathbf{1}'$ has $T$-flavor $T_f=+1$, and the representation $\mathbf{1}''$ has $T$-flavor $T_f=-1$. Since for $A_4$-triplets, the operators $S$ and $T$ do not commute, $A_4$-triplet fields cannot simultaneously have a definite $T$-flavor and a definite $S$-flavor-parity.
While the real representation of $A_4$ in Eqs.~(\ref{generator}), in which $S$ is diagonal, is useful in writing the Lagrangian, the physical meaning of our model is more transparent in the $T$-flavor representation in which $T$ is diagonal. This $T$-flavor representation is obtained from the $S$-flavor representation in (\ref{generator}) through the unitary transformation
\begin{align}
A \to A'=U^{\dag}_{\omega}AU_{\omega},
\end{align}
where $A$ is any $A_4$ matrix in the real $S$-diagonal representation and
\begin{align}
U_{\omega}=\frac{1}{\sqrt{3}}{\left(\begin{array}{ccc}
 1 &  1 &  1 \\
 1 & \omega & \omega^{2} \\
 1 & \omega^{2} & \omega
 \end{array}\right)}.
 \label{eq:Uomega}
\end{align}
In the $T$-flavor representation we have
 \begin{eqnarray}
 S'=\frac{1}{3} \, {\left(\begin{array}{ccc}
 -1 &  2 &  2 \\
 2 &  -1 & 2 \\
 2 &  2 &  -1
 \end{array}\right)}~,\qquad T'={\left(\begin{array}{ccc}
 1 &  0 &  0 \\
 0 &  \omega &  0 \\
 0 &  0 &  \omega^2
 \end{array}\right)}~.
 \label{generator2}
 \end{eqnarray}
Despite the physical advantages of the $T$-diagonal $S'$, $T'$ representation, for clarity of exposition and to avoid confusion and complications, in this paper  we use the $S$-diagonal real representation $S$, $T$ almost exclusively. For reference,
an $A_4$ triplet field with components $(a_1,a_2,a_3)$ in the $S$-diagonal real representation can be expressed in terms of $T$-flavor eigenfields
$(a_{0},a_{+1},a_{-1})$ as
\begin{align}
 a_{1} = \frac{a_{0}+a_{+1}+a_{-1}}{\sqrt{3}} , \quad
 a_{2} = \frac{a_{0}+\omega a_{+1}+\omega^2 a_{-1}}{\sqrt{3}} , \quad
 a_{3} = \frac{a_{0}+\omega^2 a_{+1}+\omega a_{-1}}{\sqrt{3}} .
 \label{eq:Ua1}
\end{align}
Inversely,
 \begin{align}
 a_{0}  = \frac{a_{1}+a_{2}+a_{3}}{\sqrt{3}} , \quad
 a_{+1}  = \frac{a_{1}+\omega^2 a_{2}+\omega a_{3}}{\sqrt{3}} , \quad
 a_{-1}  = \frac{a_{1}+\omega a_{2}+\omega^2 a_{3}}{\sqrt{3}} .
 \label{eq:Ua2}
 \end{align}
Now, in the $T$ diagonal basis the product rules of two triplets $(a_{0},a_{+1},a_{-1})$ and $(b_{0},b_{+1},b_{-1})$ according to ${\bf 3}\otimes{\bf 3}={\bf 3}_{s}\oplus{\bf
3}_{a}\oplus{\bf 1}\oplus{\bf 1}'\oplus{\bf 1}''$ are as follows
 \begin{eqnarray}
  (a_c\otimes b_c)_{{\bf 3}_{\rm s}} &=& \frac{1}{\sqrt{3}}(2a_{0}b_{0}-a_{+1}b_{-1}-a_{-1}b_{+1}, 2a_{-1}b_{-1}-a_{+1}b_{0}-a_{0}b_{+1}, 2a_{+1}b_{+1}-a_{-1}b_{0}-a_{0}b_{-1})~,\nonumber\\
  (a_c\otimes b_c)_{{\bf 3}_{\rm a}} &=& i\,(a_{-1}b_{+1}-a_{+1}b_{-1}, a_{+1}b_{0}-a_{0}b_{+1}, a_{0}b_{-1}-a_{-1}b_{0})~,\nonumber\\
  (a_c\otimes b_c)_{{\bf 1}} &=& a_{0}b_{0}+a_{+1}b_{-1}+a_{-1}b_{+1}~,\nonumber\\
  (a_c\otimes b_c)_{{\bf 1}'} &=& a_{0}b_{+1}+a_{+1}b_{0}+a_{-1}b_{-1}~,\nonumber\\
  (a_c\otimes b_c)_{{\bf 1}''} &=& a_{0}b_{-1}+a_{+1}b_{+1}+a_{-1}b_{0}~.
 \end{eqnarray}
 The $T$-flavor number of the products and sums can be easily checked by recalling that $-1-1=+1$ and $+1+1=-1$.

The connection to the geometry of the tetrahedron can be obtained if $a_0$, $a_{-1}$ and $a_{+1}$ are interpreted as spherical components of a 3-dimensional vector: $a_0=a_z$, $a_{+1}=-(a_x+ia_y)/\sqrt{2}$ and $a_{-1} = (a_x-ia_y)/\sqrt{2}$. The resulting $z$-axis joins a vertex of the tetrahedron to the center of the opposite face, $T$ is a $120^\circ$ rotation about the $z$-axis, and $S$ is a $180^\circ$ rotation about the ``diagonal'' direction $\hat{\bf x}+\hat{\bf y}+\hat{\bf z}$, which is an axis through the midpoints of two non-adjacent edges.

\section{Vacuum alignments}
When a non-Abelian discrete symmetry like our $A_4$ is considered, it is crucial to check the stability of the vacuum.
It is well know that, in the presence of two $A_{4}$ triplet Higgs fields $\chi$ and $\Phi$, Higgs potential terms involving both $\chi$ and $\Phi$ would be problematic for vacuum stability. Since the $\Theta$ and $\chi$ VEVs are very heavy, they can be decoupled from the theory at an energy scale much higher than electroweak scale. But, it is not enough for such vacuum stability to be  guaranteed. One can use extra dimensions~\cite{vacuum} to solve naturally such stability problems by separating physically between $\Theta,\chi$ and $\Phi,\eta$.
In this case, the problematic flavon-Higgs terms $V(\chi,\Phi)$ are not allowed or highly suppressed, and the scalar potential is a sum of a flavon potential $V(\Theta,\chi)$ depending only on the flavon fields and a Higgs potential $V(\eta,\Phi)$ depending only on the electroweak Higgs fields,\footnote{In Eq.~(\ref{potential}) the equal signs mean that the interactions between $\Theta,\chi$ and $\Phi, \eta$ are sufficiently small. Here ``sufficiently small" means that these interaction terms cannot ruin the imposed VEV alignment. There also needs to be a sufficiently small soft breaking term to avoid Goldstone modes resulting from the spontaneous breaking of $U(1)_X$.}
\begin{eqnarray}
V(\Theta,\chi,\Phi,\eta) &=&  V(\Theta,\chi)+V(\Phi,\eta).
\label{potential}
\end{eqnarray}
Minimization of the scalar potential is then achieved separately for the flavon and the electroweak Higgs fields. We now discuss how to realize the vacuum alignment after spontaneous flavor symmetry breaking.

\subsection{Minimization of the flavon potential}

The most general renormalizable scalar potential for the flavon fields $\Theta$ and $\chi$, invariant under $SU(2)_{L}\times U(1)_{Y}\times A_{4}\times U(1)_X\times Z_{2}$, is given by
\begin{eqnarray}
V(\Theta,\chi) &=& \mu^{2}_{\Theta}\Theta^{\ast}\Theta
+\lambda^{\Theta}\Theta^{\ast}\Theta^{\ast}\Theta\Theta~,\nonumber\\
&+& \mu^{2}_{\chi}(\chi\chi^{\ast})_{\mathbf{1}}+\lambda^{\chi}_{1}(\chi\chi)_{\mathbf{1}}(\chi^{\ast}\chi^{\ast})_{\mathbf{1}}+\tilde{\lambda}^{\chi}_{1}(\chi^{\ast}\chi)_{\mathbf{1}}(\chi^{\ast}\chi)_{\mathbf{1}}+\lambda^{\chi}_{2}
            (\chi\chi)_{\mathbf{1}^\prime}(\chi^{\ast}\chi^{\ast})_{\mathbf{1}^{\prime\prime}}\nonumber\\
&+&\tilde{\lambda}^{\chi}_{2} (\chi^{\ast}\chi)_{\mathbf{1}^\prime}(\chi^{\ast}\chi)_{\mathbf{1}^{\prime\prime}}
  +\lambda^{\chi}_{3}(\chi\chi)_{\mathbf{3}_{s}}(\chi^{\ast}\chi^{\ast})_{\mathbf{3}_{s}}+\tilde{\lambda}^{\chi}_{3}(\chi^{\ast}\chi)_{\mathbf{3}_{s}}(\chi^{\ast}\chi)_{\mathbf{3}_{s}}~\nonumber\\
&+&\lambda^{\chi\Theta}_{1}(\chi\chi^{\ast})_{\mathbf{1}}\Theta^{\ast}\Theta+\left\{\lambda^{\chi\Theta}_{2}(\chi\chi)_{\mathbf{1}}\Theta^{\ast}\Theta^{\ast}+\lambda^{\chi\Theta}_{3}(\chi\chi\chi^{\ast})_{\mathbf{1}}\Theta^{\ast}+{\rm h.c.}\right\}~,
\label{potential1}
\end{eqnarray}
Here $\mu_{\Theta}$ and $\mu_{\chi}$ have mass dimension-1, while $\lambda^{\Theta}$, $\lambda^{\chi}_{1,2,3}$, $\tilde{\lambda}^{\chi}_{1,2,3}$ are real and dimensionless and $\lambda^{\chi\Theta}_{1,2,3}$ are complex and dimensionless.

The vacuum configuration for $\chi$ and $\Theta$ is obtained by setting to zero the derivatives of $V$ with respect to each component of the scalar fields $\chi_j$ and $\Theta$. We have four minimization conditions for the four VEVs $v_{\chi_j}$ and $v_\Theta$:
 \begin{eqnarray}
  \frac{\partial V(\Theta,\chi) }{\partial \chi_{j}}\Bigg|_{\chi_j=v_{\chi_j}}=0~,\qquad\frac{\partial V(\Theta,\chi) }{\partial \Theta}\Bigg|_{\Theta=v_\Theta}=0~,\qquad{\rm for}~j=1,2,3~.
 \end{eqnarray}
Since $V(\Theta,\chi)$ is invariant under $A_4\times U(1)_X$, the space of solutions of the minimization conditions is invariant under $A_4\times U(1)_X$. Therefore it is possible to fix the phase of the VEV $\langle\Theta\rangle$ without loss of generality: we choose $v_\Theta$ real. Once an alignment of the $A_4$ triplet VEV $\langle\chi\rangle$ is chosen, the orbit of $\langle\chi\rangle$ under $A_4$ contains discrete degenerate vacua. A solution to the ensuing problem of cosmological topological defects is outside the scope of this work. We show that a minimum exists for the alignment $\langle\chi\rangle = (v_\chi, 0, 0)$. With this choice, and excluding the trivial solution where all VEVs vanish, the minimization conditions read
\begin{align}
& \Re(\lambda^{\chi\Theta}_{2} v_\chi^2) = 0, \nonumber \\
& \mu^{2}_{\chi}+2\left(\lambda^{\chi}_{1}+\tilde{\lambda}^{\chi}_{1}+\lambda^{\chi}_{2}+\tilde{\lambda}^{\chi}_{2}\right)|v_{\chi}|^2+\left(\lambda^{\chi\Theta}_{1}+2|\lambda^{\chi\Theta}_{2}|\right)v^{2}_{\Theta}= 0,\nonumber\\
&\mu^{2}_{\Theta}+\left(\lambda^{\chi\Theta}_{1}+2|\lambda^{\chi\Theta}_{2}|\right) |v_{\chi}|^2 +2\lambda^{\Theta}v^{2}_{\Theta}= 0.
 \end{align}
These have unique solution
\begin{align}
\arg v_\chi & = - \frac{1}{2} \arg \lambda^{\chi\Theta}_{2}, \nonumber \\
|v_\chi|^2 & = \frac{- \left(\lambda^{\chi\Theta}_{1}+2|\lambda^{\chi\Theta}_{2}|\right) \mu^{2}_{\Theta} + 2\lambda^{\Theta}\mu^{2}_{\chi}}{\left(\lambda^{\chi\Theta}_{1}+2|\lambda^{\chi\Theta}_{2}|\right)^2-4\left(\lambda^{\chi}_{1}+\tilde{\lambda}^{\chi}_{1}+\lambda^{\chi}_{2}+\tilde{\lambda}^{\chi}_{2}\right)\lambda^{\Theta}},\nonumber\\
v_\Theta^2 & = \frac{2\left(\lambda^{\chi}_{1}+\tilde{\lambda}^{\chi}_{1}+\lambda^{\chi}_{2}+\tilde{\lambda}^{\chi}_{2}\right) \mu^{2}_{\Theta} - \left(\lambda^{\chi\Theta}_{1}+2|\lambda^{\chi\Theta}_{2}|\right)\mu^{2}_{\chi}}{\left(\lambda^{\chi\Theta}_{1}+2|\lambda^{\chi\Theta}_{2}|\right)^2-4\left(\lambda^{\chi}_{1}+\tilde{\lambda}^{\chi}_{1}+\lambda^{\chi}_{2}+\tilde{\lambda}^{\chi}_{2}\right)\lambda^{\Theta}},
  \label{vevChi}
 \end{align}
provided the right-hand sides of the $|v_\chi|^2$ and $v_\Theta^2$ expressions are positive. It is not hard to see that the latter condition can be satisfied (for example, for $-\mu_\chi^2>0$ and $-\mu_\Theta^2>0$, and choosing $\lambda^{\Theta}$, $\lambda^{\chi}_{1,2,3}$, $\tilde{\lambda}^{\chi}_{1,2,3}$ positive to guarantee that the potential is stable at large $\Theta$ and $\chi$, small values of $\lambda^{\chi\Theta}_{1}+2|\lambda^{\chi\Theta}_{2}|$ admit physical solutions for $v_\chi$ and $v_\Theta$). Thus we impose the $\chi$ and $\Theta$ vacuum alignment
\begin{align}
\langle\chi\rangle = (v_\chi, 0, 0), \qquad \langle\Theta\rangle = v_\Theta,
\label{eq:VEVchitheta}
\end{align}
with $v_\Theta$ real.

\subsection{Minimization of the electroweak Higgs potential}
\label{sec:minSU2}
The most general renormalizable scalar potential for the electroweak Higgs fields $\eta$ and $\Phi$, invariant under $SU(2)_{L}\times U(1)_{Y}\times A_{4}\times U(1)_X\times Z_{2}$, is given by
\begin{eqnarray}
V(\eta,\Phi) &=& \mu^{2}_{\Phi}(\Phi^{\dag}\Phi)_{\mathbf{1}}+\lambda^{\Phi}_{1}(\Phi^{\dag}\Phi)_{\mathbf{1}}(\Phi^{\dag}\Phi)_{\mathbf{1}}+\lambda^{\Phi}_{2}(\Phi^{\dag}\Phi)_{\mathbf{1^\prime}}(\Phi^{\dag}\Phi)_{\mathbf{1^{\prime\prime}}}+\lambda^{\Phi}_{3}(\Phi^{\dag}\Phi)_{\mathbf{3}_{s}}(\Phi^{\dag}\Phi)_{\mathbf{3}_{s}}\nonumber\\
  &+&\lambda^{\Phi}_{4}(\Phi^{\dag}\Phi)_{\mathbf{3}_{a}}(\Phi^{\dag}\Phi)_{\mathbf{3}_{a}}+i\lambda^{\Phi}_{5}(\Phi^{\dag}\Phi)_{\mathbf{3}_{s}}(\Phi^{\dag}\Phi)_{\mathbf{3}_{a}}~\nonumber\\
 &+& \mu^{2}_{\eta}(\eta^{\dag}\eta)+\lambda^{\eta}(\eta^{\dag}\eta)^{2}~\nonumber\\
 &+& \lambda^{\eta\Phi}_{1}(\Phi^{\dag}\Phi)_{\mathbf{1}}(\eta^{\dag}\eta)
  +\lambda^{\eta\Phi}_{2}[(\eta^{\dag}\Phi)(\Phi^{\dag}\eta)]_{\mathbf{1}}+\left\{\lambda^{\eta\Phi}_{3}[(\eta^{\dag}\Phi)(\eta^{\dag}\Phi)]_{\mathbf{1}}+\text{h.c.}\right\}~.
\label{potential2}
\end{eqnarray}
Here $\mu_{\Phi}$ and $\mu_{\eta}$ have mass dimension-1, while $\lambda^{\Phi}_{1,...,5}$, $\lambda^{\eta}$ are real and dimensionless, and $\lambda^{\eta\Phi}_{1,2,3}$ are complex and dimensionless.

The $A_4$ symmetry makes $V(\eta,\Theta)$ invariant under cyclic permutations of $\Phi_1$, $\Phi_2$, and $\Phi_3$. In addition, it can be shown that the potential obtained from $V(\eta,\Phi)$ after exchanging $\Phi_1$ and $\Phi_2$ differs from the original potential by a term $-2i\lambda^{\Phi}_{5}(\Phi^{\dag}\Phi)_{\mathbf{3}_{s}}(\Phi^{\dag}\Phi)_{\mathbf{3}_{a}}$, which vanishes for real $\Phi$. Thus the potential $V(\eta,\Theta)$ is invariant under permutations of $\Phi_1$, $\Phi_2$, and $\Phi_3$ when $\Phi$ is real. It is therefore interesting to consider a CP-invariant minimum of $V(\eta,\Theta)$ symmetric under permutations of $\Phi_1$, $\Phi_2$, and $\Phi_3$, i.e., with the vacuum alignment
\begin{align}
\langle \Phi_1 \rangle = \langle \Phi_2 \rangle = \langle \Phi_3 \rangle = \frac{1}{\sqrt{2}} \begin{pmatrix} 0 \\ v_\Phi \end{pmatrix},
\qquad
\langle \eta \rangle = \frac{1}{\sqrt{2}} \begin{pmatrix} 0 \\ v_\eta \end{pmatrix},
\label{Alin}
\end{align}
with real $v_\Phi$ and $v_\eta$. Under this ansatz, and the additional assumption that $\lambda^{\eta\Phi}_{3}$ is real, the minimization conditions become
\begin{align}
\mu^{2}_{\Phi}+v^{2}_{\Phi}\left(3\lambda^{\Phi}_{1}+4\lambda^{\Phi}_{3}\right)+v_{\eta}^2\left( \frac{\lambda^{\eta\Phi}_{1}+\lambda^{\eta\Phi}_{2}}{2}+\lambda^{\eta\Phi}_{3} \right)= 0 ,
\label{vevEtaPhi1}
\\
v_{\eta}\mu^{2}_{\eta}+v_{\eta}^3\lambda^{\eta} +3v^{2}_{\Phi}v_\eta\left(\frac{\lambda^{\eta\Phi}_{1}+\lambda^{\eta\Phi}_{2}}{2}+\lambda^{\eta\Phi}_{3}\right) = 0.
\label{vevEtaPhi2}
\end{align}
We want to show that there are values of the parameters $\mu_\Phi^2$, $\mu_\eta^2$, $\lambda^{\Phi}_{1,3}$, $\lambda^{\eta}$ and $\lambda^{\eta\Phi}_{1,2,3}$ for which these two equations admit a real solution for $v_\eta$ and $v_\Phi$. For illustration, we set $v_\eta=v_\Phi$, as in our numerical work, and find a solution
\begin{align}
v_\eta = v_\Theta = \left[\frac{-\mu_\eta^2}{\lambda^\eta + \frac{3}{2}(\lambda^{\eta\Phi}_{1}+\lambda^{\eta\Phi}_{2})+3\lambda^{\eta\Phi}_{3}}\right]^{1/2}
\end{align}
provided the conditions $-\mu_\eta^2>0$, $-\mu_\Phi^2>0$, and
\begin{align}
\frac{-\mu_\eta^2}{\lambda^\eta + 3\left(\frac{\lambda^{\eta\Phi}_{1}+\lambda^{\eta\Phi}_{2}}{2}+\lambda^{\eta\Phi}_{3}\right)} = \frac{-\mu_\Phi^2}{3\lambda^{\Phi}_{1}+4\lambda^{\Phi}_{3} + \frac{\lambda^{\eta\Phi}_{1}+\lambda^{\eta\Phi}_{2}}{2}+\lambda^{\eta\Phi}_{3}}
\end{align}
are satisfied (which is possible, for example, for real and positive $\lambda^{\Phi}_{1,3}$, $\lambda^{\eta}$ and $\lambda^{\eta\Phi}_{1,2,3}$).
Hence we conclude that there are non-trivial VEV configurations for $\Phi$ and $\eta$ of the form
 \begin{eqnarray}
  \langle\Phi\rangle=\frac{v_{\Phi}}{\sqrt{2}}(1,1,1)~,\qquad\langle\eta\rangle=\frac{v_{\eta}}{\sqrt{2}}~.
  \label{Alin2}
 \end{eqnarray}

\section{}
The diagonalizing matrix $V^{f}_{L}$ ($f=\ell,\nu,d,u$) can be parameterized in terms of three mixing angles and six phases:
 \begin{eqnarray}
 V^{f}_{L}={\left(\begin{array}{ccc}
 c^{f}_{2}c^{f}_{3} &  c^{f}_{2}s^{f}_{3}e^{i\phi^{f}_{3}} &  s^{f}_{2}e^{i\phi^{f}_{2}} \\
 -c^{f}_{1}s^{f}_{3}e^{-i\phi^{f}_{3}}-s^{f}_{1}s^{f}_{2}c^{f}_{3}e^{i(\phi^{f}_{1}-\phi^{f}_{2})} &  c^{f}_{1}c^{f}_{3}-s^{f}_{1}s^{f}_{2}s^{f}_{3}e^{i(\phi^{f}_{1}-\phi^{f}_{2}+\phi^{f}_{3})} &  s^{f}_{1}c^{f}_{2}e^{i\phi^{f}_{1}} \\
 s^{f}_{1}s^{f}_{3}e^{-i(\phi^{f}_{1}+\phi^{f}_{3})}-c^{f}_{1}s^{f}_{2}c^{f}_{3}e^{-i\phi^{f}_{2}} &  -s^{f}_{1}c^{f}_{3}e^{-i\phi^{f}_{1}}-c^{f}_{1}s^{f}_{2}s^{f}_{3}e^{i(\phi^{f}_{3}-\phi^{f}_{2})} &  c^{f}_{1}c^{f}_{2}
 \end{array}\right)}Q_{f}~,
 \label{Vl}
 \end{eqnarray}
where $s^{f}_{i}\equiv \sin\theta^{f}_{i}$ and $c^{f}_{i}\equiv \cos\theta^{f}_{i}$. The diagonal phase matrix $Q_{f}=\diag(e^{i\xi_{1}},e^{i\xi_{2}},e^{i\xi_{3}})$ can be rotated away by a phase redefinition of the fermion fields.

The parameters induced by higher dimensional operators, appearing in Eq.~(\ref{meff}), are defined as
 \begin{align}
 &
 \delta_{11}=4 F x^{s}_{1}\,,
 \nonumber\\
 &
 \delta_{22}= (F-3G)x^{s}_{2}y_2+i\sqrt{3}(F+G)x^{a}_{2}y_2\,,
 \nonumber\\
 &
 \delta_{33}=(F-3G)x^{s}_{3}y_3-i\sqrt{3}(F+G)x^{a}_{3}y_3\,,
 \nonumber\\
 &
 \delta_{12}= -F (x^{s}_{2}+x^{s}_{1} y_2)-i\sqrt{3} (Fx^{a}_{2}+G x^{a}_{1} y_2)\,,
 \nonumber\\
 &
 \delta_{13}=-F (x^{s}_{3}+ x^{s}_{1} y_3) +i\sqrt{3} (Fx^{a}_{3}+G x^{a}_{1} y_3)\,,
 \nonumber\\
 &
 \delta_{23}=\tfrac{F+3G}{2}(x^{s}_{2}y_3+x^{s}_{3}y_2)+i\sqrt{3}\tfrac{F-G}{2}(x^{a}_{2}y_3-x^{a}_{3}y_2)\,,\label{C1}
 \end{align}
and
\begin{align}
 &
 \gamma_{11}=2 F (x^{s}_{1})^2 - 2 G (x^{a}_{1})^2 \,,
 \nonumber\\
 &
 \gamma_{22}=-\tfrac{G+3F}{2}(x^a_2)^2+\tfrac{F+3G}{2}(x^s_2)^2+i\sqrt{3}(F-G)x^{a}_{2}x^{s}_{2}\,,
 \nonumber\\
 &
 \gamma_{33}=-\tfrac{G+3F}{2}(x^a_3)^2+\tfrac{F+3G}{2}(x^s_2)^2-i\sqrt{3}(F-G)x^{a}_{3}x^{s}_{3}\,,
 \nonumber\\
 &
 \gamma_{12}= -F x^{s}_{1} x^{s}_{2}+Gx^{a}_{1} x^{a}_{2}-i\sqrt{3} (Fx^{a}_{2}x^{s}_{1}-G x^{a}_{1} x^{s}_{2})\,,
 \nonumber\\
 &
 \gamma_{13}= -F x^{s}_{1} x^{s}_{3}+Gx^{a}_{1} x^{a}_{3}+i\sqrt{3} (Fx^{a}_{3}x^{s}_{1}-G x^{a}_{1} x^{s}_{3})\,,
 \nonumber\\
 &
 \gamma_{23}=\tfrac{3F-G}{2}x^{a}_{2}x^{a}_{3}+\tfrac{F-3G}{2}x^{s}_{2}x^{s}_{3}+i\sqrt{3}\tfrac{F+G}{2}(x^{a}_{2}x^{s}_{3}-x^{a}_{3}x^{s}_{2})\,,\label{C2}
 \end{align}
with
 \begin{eqnarray}
  x^{s(a)}_{i}\equiv \frac{\hat{x}^{s(a)}_{i}}{\hat{y}^{\nu}_{1}}\,.
 \end{eqnarray}

\acknowledgments{
PG was supported in part by NSF grant PHY-1068111 at the University of Utah.
}


\end{document}